%% file: main.tex
\documentclass[conference]{IEEEtran}
\IEEEoverridecommandlockouts
\usepackage{amsmath,amsfonts}
\usepackage{graphicx}
\usepackage{textcomp}
\usepackage{xcolor}
\usepackage{multirow}
\usepackage{hhline}
\usepackage{pifont}
\usepackage{threeparttable}
\usepackage{color}
\usepackage{amsthm}
\usepackage{amssymb}
\usepackage{amsmath,stackengine}
\usepackage{mathrsfs}
\theoremstyle{definition}
\newtheorem{dfn}{Definition}

\usepackage{subcaption}
\usepackage{array}
\usepackage{algorithm}
\usepackage[noEnd]{algpseudocodex}
\usepackage{booktabs}
\usepackage{tcolorbox}
\usepackage{enumitem}
\usepackage{varwidth}
\usepackage{hyperref}
\usepackage{xspace}
\usepackage{multicol}
\usepackage{adjustbox}
\usepackage[dvipsnames]{xcolor}
\usepackage{makecell}
\usepackage{rotating}
\usepackage{newtxtext,newtxmath}
\usepackage{setspace}

\algrenewcommand\algorithmicindent{0.6em}

\newcommand{\draftparagraph}[1]{\noindent{\textit{\textbf {#1. }}}}

\newcommand{\sample}{\xleftarrow{\$}}

\def\ifeq{\mathrel{\ooalign{%
  \raisebox{0.85\height}{$\hskip0.25ex\scriptstyle?$}\cr$\raisebox{-0.15\height}{=}$\cr}}}

\def\Assert{\textbf{assert}\xspace}
\def\fdsid{h.\vspace{0.2ex}\hspace{0.2ex}\belowbaseline[-2.15ex]{\framebox(1.6em,2.7ex)[c]{\rule{0pt}{1.2ex}$sid$}}\xspace}

\def\fdidx{h.\vspace{0.2ex}\hspace{0.2ex}\belowbaseline[-2.2ex]{\framebox(1.6em,2.7ex)[c]{\rule{0pt}{1.2ex}$idx$}}\xspace}

\def\fdphase{h.\vspace{0.2ex}\hspace{0.2ex}\belowbaseline[-1.9ex]{\framebox(2.9em,2.7ex)[c]{\rule{0pt}{1.2ex}$phase$}}\xspace}
\def\fddir{h.\vspace{0.2ex}\hspace{0.2ex}\belowbaseline[-2.1ex]{\framebox(1.6em,2.7ex)[c]{\rule{0pt}{1.2ex}$dir$}}\xspace}

\def\fdtau{h.\vspace{0.2ex}\hspace{0.2ex}\belowbaseline[-1.9ex]{\framebox(0.7em,2.7ex)[c]{\rule{0pt}{1.2ex}$\tau$}}\xspace}
\def\fdrho{h.\vspace{0.2ex}\hspace{0.2ex}\belowbaseline[-1.8ex]{\framebox(0.7em,2.7ex)[c]{\rule{0pt}{1.2ex}$\rho$}}\xspace}

\def\fdE{h.\vspace{0.2ex}\hspace{0.2ex}\belowbaseline[-2.1ex]{\framebox(0.9em,2.7ex)[c]{\rule{0pt}{1.2ex}$E$}}\xspace}
\def\fdC{h.\vspace{0.2ex}\hspace{0.2ex}\belowbaseline[-2.1ex]{\framebox(0.9em,2.7ex)[c]{\rule{0pt}{1.2ex}$C$}}\xspace}
\def\fdPi{h.\vspace{0.2ex}\hspace{0.2ex}\belowbaseline[-2.1ex]{\framebox(0.9em,2.7ex)[c]{\rule{0pt}{1.2ex}$\Pi$}}\xspace}
\def\fdK{h.\vspace{0.2ex}\hspace{0.2ex}\belowbaseline[-2.1ex]{\framebox(0.9em,2.7ex)[c]{\rule{0pt}{1.2ex}$K$}}\xspace}
\def\fdM{h.\vspace{0.2ex}\hspace{0.2ex}\belowbaseline[-2.1ex]{\framebox(1.1em,2.7ex)[c]{\rule{0pt}{1.2ex}$M$}}\xspace}

\def\fdsrc{h.\vspace{0.2ex}\hspace{0.2ex}\belowbaseline[-2.0ex]{\framebox(1.5em,2.7ex)[c]{\rule{0pt}{1.2ex}$src$}}\xspace}
\def\fddst{h.\vspace{0.2ex}\hspace{0.2ex}\belowbaseline[-2.2ex]{\framebox(1.7em,2.7ex)[c]{\rule{0pt}{1.2ex}$dst$}}\xspace}

\newtheorem{prop}{Proposition}[section]
\newtheorem{thm}{Theorem}[section]
\newenvironment{proofsketch}{
  \proof}{\endproof}

\def\pname{\textsc{Lacan}\xspace}

\begin{document}

\title{\pname: Making Accountability in Anonymous Networks Real
\thanks{This work was supported by JSPS KAKENHI Grant Numbers 23K28073 and 24KJ1629.}}

\author{
\IEEEauthorblockN{Naoya Takada\IEEEauthorrefmark{1}, 
Yutaro Yoshinaka\IEEEauthorrefmark{1}, 
Kentaro Kita\IEEEauthorrefmark{1}, 
Junji Takemasa\IEEEauthorrefmark{1},
Yuki Koizumi\IEEEauthorrefmark{1}, and 
Toru Hasegawa\IEEEauthorrefmark{2}}
\IEEEauthorblockA{\IEEEauthorrefmark{1}Graduate School of Information Science and Technology, The University of Osaka}
\IEEEauthorblockA{\IEEEauthorrefmark{2}Faculty of Materials for Energy, Shimane University}
}

\maketitle

\input{abstract}


\input{introduction}
\input{motivation}

\input{preliminaries}
\input{rationale}

\input{protocol}
\input{security}

\input{evaluation}

\input{discussion}
\input{conclusion}

\newpage
\bibliographystyle{IEEEtran}
\bibliography{reference}

\end{document}

%% file: abstract.tex
\begin{abstract}
\emph{Anonymity} and \emph{accountability} are essential properties for our everyday activity on the Internet. However, they appear contradictory, and their reconciliation remains far from \emph{reality}. Existing approaches fall short in this regard, as they either rely on an on-path trustee, per-packet authorization, per-packet public-key cryptography, or per-session intervention by a central authority.
We propose \textsc{Lacan}, a protocol that reconciles anonymity and accountability within a realistic design. In \textsc{Lacan}, a sender enjoys anonymity provided by on-path relays, as long as she complies with a contract established with the receiver. Upon a contract violation, the verifier, an off-path trustee on behalf of the receiver, links the malicious message to the sender's identity \emph{indirectly} via the packet, path, and session, thereby reducing public-key operations from per-packet to per-session. This linkage remains robust even against malicious relays and receivers, grounded in our novel \emph{chain of successor proofs} for accountable path reconstruction, together with traceable signatures, path validation, and key-committing encryption. We analyze the anonymity and accountability, implement the protocol, and evaluate the performance. \looseness=-1
\end{abstract}

%% file: introduction.tex
\section{Introduction}

\emph{Anonymity}, unlinking the actor's identity from her actions, is undoubtedly an essential right for us citizens, especially given that privacy on the Internet is under attack from powers including national authorities and corporate capital~\cite{FARRELL2014}.

However, anonymity sometimes works to our disadvantage by allowing attackers to stay undercover and evade rightful sanctions. In this regard, we desire \emph{accountability}, the ability to link any malicious online action to its actor's identity.

This paper presents a reconciliation of the seemingly contradictory notions of anonymity and accountability on the Internet.
In particular, we focus on the \emph{reality} of protocol design in both the security and performance perspectives, which prior proposals~\cite{naylor2014balancing, lee2016source, jonathancontractual, kopsell2006revocable} fall short in this regard:
Trusting a single on-path node~\cite{lee2016source} undermines the virtue of distributed anonymous communication.
Per-packet expensive operations, such as briefing an off-path trustee~\cite{naylor2014balancing} and public-key cryptography~\cite{jonathancontractual, kopsell2006revocable}, significantly degrade performance.

In our proposed protocol, named \pname, enables the realistic coexistence of anonymity and accountability based on \emph{contract}~\cite{jonathancontractual}, under the realistic threat model. 
Our model is a hybrid of the \emph{local adversary} assumption~\cite{hsiao2012lap,sankey2014dovetail,chen2017phi} for anonymity and the \emph{honest majority} assumption for accountability.
The communicating \emph{sender} and \emph{receiver} agree in advance on a non-retroactive contract, under which the sender enjoys anonymity as long as she adheres to it. 
Only if the contract is violated, the receiver can link the message plaintext to the sender's identity with the assistance of an off-path trustee, \emph{verifier}. 
Since computations for accountability are performed only on a per-session basis, the protocol achieves near-ideal performance, given that each session carries a sufficiently large amount of data and that contract violations occur infrequently.

This linkage is achieved in four stages---linking the session to the sender's identity, the communication path to the session, the packet to the path, and the message plaintext to the packet---thereby reducing signature computation from per-packet to per-session. 
First, the session is linked to the sender's identity by the verifier, who can open a per-session \emph{traceable signature}~\cite{kiayias2004traceable}. 
Second, the path is linked to the session as the verifier backtraces and reconstructs the path.
This backtracing procedure is secured against dishonest relays through a \emph{chain of successor proofs}, which serves as a commitment to truthful reconstruction of a path once the contract is violated.
Third, the packet is linked to the path by the verifier, who queries each on-path relay to confirm whether it has forwarded the packet.
If a majority of relays respond affirmatively, the verifier concludes that the packet indeed traversed the path.
Finally, the verifier links the message plaintext to the packet by confirming successful decryption of the packet using \emph{key-committing authenticated encryption (AE)}~\cite{farshim2017security,albertini2022abuse}, which guarantees unique decryption.

Our main contributions are summarized as follows:
\setlength{\leftmargini}{1.2em}
\begin{itemize}[topsep=0.35mm]
    \item A realistic threat model that enables efficient reconciliation of anonymity and accountability (Section~\ref{sec:threat_model})
    \item Four-stage linking of a message to its sender, which reduces signatures from per-packet to per-session (Section~\ref{sec:intuition})
    \item A chain of successor proofs, which convince the receiver that the path can be reconstructed by the verifier (Section~\ref{sec:link2})
    \item \pname, a proposed protocol, and its security analysis, covering attacks on anonymity and accountability (Sections~\ref{sec:protocol},\ref{sec:security})
    \item Implementation of \pname and its evaluation (Section~\ref{sec:perf})
\end{itemize}

%% file: motivation.tex
\section{Motives and Goals}
\label{sec:motivation}

\subsection{Motivating Scenarios}
\draftparagraph{Mass surveillance}
Consider the online activity of an ordinary and privacy-conscious user. 
She desires \emph{anonymity}---to conceal the fact that she visits specific websites---because most network parties are incentivized to expose this fact and use it to infer, for example, her political spectrum and products she may be interested in, which goes against her will.

\vspace{0.5mm}
\draftparagraph{Abusive users}
Meanwhile, malicious users may, for example, launch cyber attacks or spread discriminatory speech, using anonymity to conceal their identities and evade rightful sanction.
We focus on abuse-prone websites and services, such as social media, file hosting, and web APIs. 
Such websites and services demand \emph{accountability}---in other words, to identify malicious users and protect themselves from future aggression.

\vspace{0.5mm}
\draftparagraph{Uncooperative network}
However, victimized websites generally cannot rely on network parties to ensure accountability, because disclosing user information may be neither justifiable nor beneficial.
Disclosure policies differ across ISPs~\cite{soghoian2011end}, which often have adversarial relationships with law enforcement~\cite{vermeer2018identifying}. 
Cross-border disclosure, common in forensics in anonymous networks, further requires case-by-case decisions~\cite{ebert2024responding}.
Worse, network parties may collude with malicious senders to attack websites, or collude with websites to frame innocent senders.

\vspace{0.5mm}
A property we refer to as \emph{reality} is grounded in mutual distrust among network parties, all of whom may behave maliciously.
These parties must be technically enforced and encouraged cooperate honestly, regardless of their societal relationships.

\subsection{Contract}
\label{sec:contract}
We define the malicious behavior of a sender as a violation of the \emph{contract}, inspired by the proposal in RECAP~\cite{jonathancontractual}.

A contract is a \emph{well-defined}, \emph{deterministic}, and \emph{universally and polynomial-time computable} boolean function over message plaintexts, $f: \{0, 1\}^* \rightarrow \{0, 1\}$, whose range indicates whether the message complies with the contract. 
An example of a contract is checking whether a message contains no word from a public, predefined blocklist.
Conversely, ambiguous, non-deterministic, dependent on a secret, or inefficient functions are not contracts.
A broader definition of contracts with more expressiveness will be discussed in Section~\ref{sec:discussion}.

Each receiver defines a contract specifying which messages they are willing to accept and publishes it to the \emph{public directory}.
A sender wishing to communicate with a receiver retrieves the contract and initiates communication only if it is acceptable.
If the receiver detects a message that violates the contract, it initiates a process to hold the sender accountable.

Receivers may modify their contract at any time.
Since all modifications are recorded in the public directory, any receiver $R$'s contract at any timestamp $ts \in \mathbb{N}$ is publicly evaluable.
We denote this contract by $\mathscr{C}_R: \mathbb{N} \times \{0,1\}^* \rightarrow \{0,1\}$.
A receiver cannot retroactively accuse a sender under a modified contract. 
Each communication is bound to the contract that is in effect when the communication is initiated. Thus, a sender needs to retrieve the latest contract only when initiating communication with a receiver, rather than before transmitting every packet. \looseness=-1

This paper considers only messages carried by a single packet, although it naturally extends to those spanning multiple packets.

\subsection{Anonymity Goals}
\draftparagraph{Relationship anonymity} 
Hide \emph{who communicates with whom}. 
Precisely, the identities of the sender and receiver are unlinkable for any adversary~\cite{pfitzmann2001anonymity,backes2013anoa,kuhn2018privacy}. 
Because this notion is implied by either \emph{sender anonymity} or \emph{receiver anonymity}, it applies even when a compromised party is in a position that inevitably learns the sender (e.g., the first-hop relay) or the receiver (e.g., the receiver itself).
It is a common objective in the design and analysis of anonymous communication protocols~\cite{shmatikov2006measuring,feigenbaum2012probabilistic,sankey2014dovetail}. 

\vspace{0.5mm}
\draftparagraph{Session unlinkability} 
Hide \emph{whether the senders of any given two sessions are the same}~\cite{kuhn2018privacy} from any adversary, unless the sender of either session has previously violated the contract.

\subsection{Accountability Goals}
\draftparagraph{Sender accountability} 
Once the protocol succeeds, the receiver can identify the sender of any message plaintext that violates the contract.
In other words, the protocol revokes the anonymity of the malicious sender, and thus also relationship anonymity.

\vspace{0.5mm}
\draftparagraph{Sender traceability} 
Once the protocol succeeds, the receiver can link any two sessions established by the same sender who previously violated the contract.
In other words, the protocol revokes the session unlinkability of the malicious sender, allowing the receiver to shut off future session establishment attempts. \looseness=-1

\subsection{Reality Goals}
\draftparagraph{Participant accountability} 
When the protocol fails, the trustee can identify the party whose misbehavior causes the failure.
Detected misbehavior includes relays that lie or refuse to cooperate, and receivers that falsely report a contract violation. 

\vspace{0.5mm}
\draftparagraph{Dataplane efficiency}
The sender, receiver, and relays are not required to perform expensive operations, such as per-packet public-key cryptography or per-packet/per-session trustee intervention.
Honest participation in the protocol remains reasonable.

\subsection{Related Work and Our Novelty}
\draftparagraph{Anonymity}
Anonymous communication protocols can be classified by whether they provide \emph{bitwise unlinkability}~\cite{danezis2003mixminion}.
Bitwise unlinkability means that, at any relay, an outgoing packet is computationally unlinkable from its corresponding incoming packet.
This property ensures that the identical packet appears entirely different at any two distinct points along the path, preventing colluding adversaries from tracing the path.

\emph{Mix networks}~\cite{chaum1981untraceable,danezis2003mixminion} and \emph{Onion Routing}~\cite{goldschlag1996hiding,dingledine2004tor} provide bitwise unlinkability by recursively decrypting packets at each relay, ensuring anonymity even against colluding adversaries.
In contrast, \emph{lightweight anonymity protocols}~\cite{hsiao2012lap,sankey2014dovetail,chen2017phi}, where relays do not modify packet payloads, lack bitwise unlinkability and thus provide anonymity only under the \emph{local adversary} assumption, wherein at most one on-path node is corrupted.

These protocols provide unconditional anonymity within their threat models and cannot revoke the anonymity of abusive users.

\vspace{0.5mm}
\draftparagraph{Accountability}
AIP~\cite{andersen2008accountable} and its successors~\cite{mirkovic2008building,yang2009internet} incorporate accountability into the Internet.
They rely on an on-path trustee, such as a router or an ISP, to inspect packet provenance and filter malicious traffic passing through the path.

However, since these designs do not target anonymity, they expose both sender and receiver identities.
In particular, authorization by an on-path node leaks sender-related information to verifying nodes.
Moreover, trusting a single on-path node conflicts with the principle of anonymous communication, as that node can de-anonymize or frame the sender.

\newcommand{\tabpack}[1]{{\renewcommand{\arraystretch}{0.5}\begin{tabular}{@{}c@{}} #1 \end{tabular}}}
\def\tabvspace{0.5ex}

\newcommand{\cmark}{{\small \textcolor{Green}{\ding{51}}}\xspace}
\newcommand{\xmark}{{\small \textcolor{OrangeRed}{\ding{55}}}\xspace}
\newcommand{\bhyphen}{{\small \rule[0.5ex]{0.55em}{0.25ex}}\hspace{0.2ex} \xspace}

\begin{table*}[t]
    \caption{Comparison of \pname with existing protocols}
    
    \label{tab:related}
    \scriptsize
    \centering
    \begin{tabular}{@{\hskip2.5mm}c@{\hskip2mm}c@{\hskip2mm}c@{\hskip3mm}c@{\hskip1.5mm}c@{\hskip2mm}c@{\hskip1.5mm}c@{\hskip2mm}c@{\hskip1.5mm}c@{\hskip2mm}}
    \toprule
    \noalign{\vskip -0.5ex}
    & & & \multicolumn{2}{c}{\textbf{Anonymity}} & \multicolumn{2}{c}{\textbf{Accountability}} & \multicolumn{2}{c}{\textbf{Reality}} \\
    \noalign{\vskip -0.5ex}
    \cmidrule(l{0.0mm}r{2.0mm}){4-5} \cmidrule(l{0mm}r{2.0mm}){6-7} \cmidrule(l{0mm}r{2.0mm}){8-9}
    \noalign{\vskip 0.0ex}
    & \tabpack{ \\ Protocol} & \tabpack{ \\ Technique} & \tabpack{Relationship \\ anonymity} & \tabpack{Bitwise \\ unlinkability} & \tabpack{Trustee for \\ sender accountability} & \tabpack{Sender \\ traceability} & \tabpack{Participant \\ accountability} & \tabpack{Dataplane \\ efficiency} \\
    \noalign{\vskip -0.2ex}
    \midrule
    \noalign{\vskip -0.2ex}
    \multirow{3}{*}[0.0mm]{\rotatebox[origin=c]{90}{\tabpack{Light-\\weight}}}
    & APIP~\cite{naylor2014balancing} & \tabpack{Authorization \\ by trustee} & \tabpack{\xmark (revealed  \\ at 1st hop)} & \xmark & \tabpack{\cmark \textcolor{Green}{\textbf{Off-path}} \\ (accountability delegate)} & \tabpack{\cmark \\ (by verifier)} & \tabpack{\xmark \\ (verifier, receiver)} & \tabpack{\xmark~(per-packet \\ briefing to trustee)} \\
    \noalign{\vskip \tabvspace}
    & APNA~\cite{lee2016source} & \tabpack{Authorization \\ by trustee} & \tabpack{\xmark (anonymity  \\ up to ISP level)} & \xmark & \tabpack{\xmark \textcolor{OrangeRed}{\textbf{On-path}} \\ (sender's ISP)} & \tabpack{\cmark \\ (by sender's ISP)} & \tabpack{\xmark \\ (receiver)} & \cmark \\
    \noalign{\vskip -0.2ex}
    \midrule
    \noalign{\vskip -0.2ex}
    \multirow{2}{*}[0.9mm]{\rotatebox[origin=c]{90}{\tabpack{Patching OR \\ circuit creation}}}
    & \tabpack{Diaz and \\ Preneel~\cite{diaz2007accountable}} & \tabpack{Verifiable encryption \\ of sender's pseudonym} & \cmark & \cmark & \tabpack{\cmark \textcolor{Green}{\textbf{Off-path}} \\ (judge)} & \tabpack{\cmark \\ (by relays)} & \tabpack{\xmark \\ (receiver)} & \tabpack{\bhyphen~(dataplane is \\ not specified)} \\
    \noalign{\vskip \tabvspace}
    & BackRef~\cite{backes2014backref} & \tabpack{Chain of signatures \\ computed by relays} & \cmark & \cmark & \tabpack{\cmark \textcolor{Green}{\textbf{Off-path}} \\ (verifier)} & \xmark &  \tabpack{\xmark \\ (relays, receiver)} & \tabpack{\bhyphen~(dataplane is \\ not specified)} \\
    \noalign{\vskip \tabvspace}
    & A-Tor~\cite{cai2017ator}  & \tabpack{Chain of signatures \\ computed by relays} & \cmark & \cmark & \tabpack{\cmark \textcolor{Green}{\textbf{Off-path}} \\ (auditor)} & \xmark & \tabpack{\xmark \\ (receiver)} & \tabpack{\bhyphen~(dataplane is \\ not specified)} \\
    \noalign{\vskip -0.2ex}
    \midrule
    \noalign{\vskip -0.2ex}
    \multirow{4}{*}[1.0mm]{\rotatebox[origin=c]{90}{\tabpack{Per-packet proofs + \\ Onion Routing (OR)}}}
    & THEMIS~\cite{xu2012accountable} & \tabpack{Proxy re-encryption \\ of packets} & \cmark & \cmark & \tabpack{\cmark \textcolor{Green}{\textbf{Off-path}} \\ (law enforcement agency)} & \xmark & \cmark & \tabpack{\xmark~(per-packet \\ re-encryption)} \\
    \noalign{\vskip \tabvspace}
    & \tabpack{K{\"o}psell \\ et al.~\cite{kopsell2006revocable}} & \tabpack{$(t_N, n)$-threshold \\ group signatures} & \cmark & \cmark & \tabpack{\cmark \textcolor{Green}{\textbf{Off-path}} \\ (law enforcement agency)} & \xmark & \tabpack{\cmark~(under honest \\ $t_N$ out of $n$ relays)} & \tabpack{\xmark~(per-packet \\ group signatures)} \\
    \noalign{\vskip \tabvspace}
    & RECAP~\cite{jonathancontractual} & \tabpack{Contract, \\ group signatures}& \cmark & \cmark & \tabpack{\cmark \textcolor{Green}{\textbf{Off-path}} \\ (accountability server)} & \xmark & \cmark & \tabpack{\xmark~(per-packet \\ group signatures)} \\
    \noalign{\vskip \tabvspace}
    & APGS~\cite{xia2020apgs} & \tabpack{Group signatures, \\ challenge caching} & \cmark & \cmark & \tabpack{\xmark \textcolor{OrangeRed}{\textbf{On-path}} \\ (sender's ISP)} & \tabpack{\cmark \\ (by sender's ISP)} & \tabpack{\xmark\\(relays)} &  \tabpack{\xmark (per-packet \\ group signatures)} \\
    \noalign{\vskip -0.2ex}    
    \midrule
    \noalign{\vskip -0.2ex}
    & \textbf{\small \pname} & \tabpack{Four-stage linking of \\ plaintext to sender} & \cmark & \xmark & \tabpack{\cmark \textcolor{Green}{\textbf{Off-path}} \\ (verifier)} & \tabpack{\cmark \\ (by receiver)} & \tabpack{\cmark~(under honest \\ majority of relays)} & \cmark \\
    \noalign{\vskip -0.2ex}
    \bottomrule
    \end{tabular}
\end{table*}

\vspace{0.5mm}
\draftparagraph{Reconciliation of them}
Table~\ref{tab:related} indicates that existing approaches fall short of reconciling anonymity and accountability in realistic designs.
They can be grouped into three lines.

The first line enhances AIP with privacy protection.
It designs lightweight protocols in which a trustee provides pseudonym-based anonymity and per-packet authorization for accountability, but these protocols remain insufficient in both anonymity and reality.
APIP~\cite{naylor2014balancing} relies entirely on sender pseudonyms for anonymity and thus cannot ensure relationship anonymity at the first-hop router. 
Packet verification and reporting are delegated entirely to routers and receivers, respectively, leaving their accountability unguaranteed.
Sender accountability also requires per-packet briefing to the trustee, incurring substantial communication overhead.
APNA~\cite{lee2016source} eliminates this briefing by trusting the sender's ISP; however, such an assumption conflicts with relationship anonymity.
Also, it cannot provide the sender and receiver with anonymity sets beyond their ISPs. 

The second line modifies per-session procedures in Onion Routing so that a trustee can later reconstruct the path when the receiver reports a session~\cite{diaz2007accountable,backes2014backref,cai2017ator}.
However, these protocols allow the receiver to report an entire session without identifying a suspected packet, leaving the sender vulnerable to framing.
BackRef~\cite{backes2014backref} is also vulnerable to attacks where a malicious relay shifts to an innocent sender the responsibility claimed by a receiver on a malicious path.

The third line adds per-packet proofs to Onion Routing, allowing a trustee to identify the packet sender despite malicious relays or receivers~\cite{xu2012accountable,kopsell2006revocable,jonathancontractual}. 
Those proofs rely on either proxy re-encryption or group signatures, both requiring per-packet public-key cryptography.
APGS~\cite{xia2020apgs} reduces per-packet computation at the receiver by having other relays verify group signatures.
However, it still requires per-packet signatures and is vulnerable to the sender colluding with a relay.

Our approach departs from all three lines of prior work.
\pname guarantees relationship anonymity and provides sender and participant accountability through an off-path trustee, without per-packet public-key cryptography and authorization.
Unlike prior protocols that treat accountability as direct linking, we formalize it as a composition of orthogonal cryptographic links. 
We decompose the end-to-end linkage between a malicious plaintext and its sender into separate links across a session, path, and packet. 
This decomposition eliminates per-packet signatures while preserving unforgeability and non-repudiation.
The only unavoidable trade-off is the loss of bitwise unlinkability.
Allowing on-path nodes to verify packet traversal without per-packet signatures requires that they recognize identical packets. 
Thus, an adversary observing multiple on-path locations may correlate packets, as in lightweight anonymity protocols.

%% file: preliminaries.tex
\section{Preliminaries}

\subsection{System Model}
The system consists of \emph{senders}, \emph{receivers}, \emph{relays}, a \emph{verifier}, and a \emph{public directory}, all connected via a network.

$\mathcal{U}$ denotes the set of all parties in the network.
The sender $S$ is a user who initiates the communication. 
The receiver $R$ is the user who is designated by the sender as the peer and who accepts the communication.
The relays $N_i$ are parties designated by the sender to mediate the communication. 
When $S$ selects $n$ relays, the resulting path is denoted by $(S, N_1, ..., N_n, R)$. 
$N_0$ and $N_{n+1}$ are used interchangeably with $S$ and $R$, respectively.

As in widely deployed systems such as Tor~\cite{dingledine2004tor} and I2P~\cite{jrandom2003invisible}, we assume that relays run on computers voluntarily. 
We adopt the \emph{integrated-system model}~\cite{scherer2024provable}, in which the receiver directly participates in the protocol.
This differs from Tor's \emph{service model}, where the last relay terminates the protocol and establishes a TLS connection to the receiver on behalf of the sender.  \looseness=-1

The verifier $V$ and the public directory $D$ are unique central parties whose identities are known to all participants.
The difference between them is that the former performs arbitrary computations, whereas the latter serves only as a bulletin board, accepting submissions from any party and publishing \emph{which party has submitted what values so far}, typically realized using mechanisms such as a public key infrastructure (PKI).

\subsection{Threat Model}
\label{sec:threat_model}
We assume \emph{active}, \emph{polynomial-time}, and \emph{bitwise} adversaries with a limited compromising capability.

Active adversaries not only eavesdrop but also alter, redirect, inject, and drop messages to extract information from honest parties and interfere with their interactions. 
Polynomial-time adversaries are those whose allowable algorithms have computational complexity bounded by a polynomial in the security parameter.
Bitwise adversaries cannot obtain timing or side-channel information, which will be discussed in Section~\ref{sec:discussion}.

The compromising capability is as follows: 
(1) The verifier and the public directory are trusted;
(2a) At most one node on a path selected by an honest sender is corrupted;
(2b) A majority of the on-path relays must be honest even if the sender is malicious; and
(3) Off-path parties may be compromised in an arbitrary number.
Formally, the set of compromised parties $\mathcal{A} \subset \mathcal{U}$ satisfies, $
    \underline{ \left(V,D \notin \mathcal{A} \right) \hspace{-0.1ex}}_\text{(1)}\hspace{-0.6ex} \land \bigl( \underline{ \left( \left| \mathcal{A} \cap \left\{N_1, ..., N_n, R\right\} \right| \leq 1 \right) \hspace{-0.1ex}}_\text{(2a)}\hspace{-1.0ex} \lor{}$ $
    \underline{ \left( \left( S \in \mathcal{A} \right)  \land \left( \left| \mathcal{A} \cap \left\{N_1, ..., N_n\right\} \right| < n/2 \right) \right) \hspace{-0.1ex}}_\text{(2b)}\hspace{-0.1ex} \bigr)$.
Case (2a), where the sender is honest, corresponds to the condition under which anonymity is guaranteed.
This setting follows the \emph{local adversary} assumption, wherein no anonymity is guaranteed against colluding adversaries as a consequence of foregoing bitwise unlinkability.
Case (2b), where the sender is malicious, corresponds to the condition required for accountability to hold, namely the \emph{honest majority} assumption.
Under this assumption, voting by the $n$ relays reaches the correct consensus regarding the traversal of a given packet, thereby eliminating the need for per-packet authorization or public-key cryptographic operations.

Relaxing the trust in the verifier is discussed in Section~\ref{sec:discussion}. 
Assuming a trusted public directory is common~\cite{canetti2016universally}, and it can also be distributed to parties with an honest majority~\cite{dykcik2018blockpki}. \looseness=-1

\subsection{Cryptographic Foundations}
\subsubsection{Basic Primitives}
$(\mathsf{Enc}, \mathsf{Dec})$ denotes an IND-CCA2 encryption scheme such that, for every key $k$ and message $m$, $\mathsf{Dec}(k, \mathsf{Enc}(k, m)) = m$ except with negligible probability (ewnp.). 
$\mathsf{MAC}$ is an EUF-CMA message authentication code (MAC) scheme with a canonical verification algorithm.
We omit the key-generation algorithms of these primitives, assuming that they are executed properly in a separate process.
$(\mathsf{Gen}, \mathsf{Sign}, \mathsf{Verify})$ is an EUF-CMA signature scheme such that, for every key pair $(sk, pk) \leftarrow \mathsf{Gen}(1^\kappa)$ and message $m$, $\mathsf{Verify}(pk, m, \mathsf{Sign}(sk, m)) = 1$ ewnp. 
$\mathsf{Com}$ is a commitment scheme with a canonical open phase, taking a $\kappa$-bit randomness $r$ and message $m$ as input.
$\mathsf{H}$ is a cryptographic hash function.

\subsubsection{Traceable Signature}
A 7-tuple of algorithms ($\mathsf{TSetup}$, $\mathsf{TJoin}$, $\mathsf{TSign}$, $\mathsf{TVerify}$, $\mathsf{TOpen}$, $\mathsf{TReveal}$, $\mathsf{TTrace}$) defines a \emph{traceable signature scheme} in the random oracle model~\cite{kiayias2004traceable}. 
The trusted \emph{group manager (GM)} first runs $\mathsf{TSetup}$ to generate a \emph{group secret key} $gsk$ and a \emph{group public key} $gpk$. 
A new user joins the group by running $\mathsf{TJoin}$ with the GM to obtain a \emph{member secret key} $msk$.
$\mathsf{TSign}$ takes $msk$ and a message $m$ and outputs a signature $\sigma$. 
$\mathsf{TVerify}$ takes $gpk$, $\sigma$, $m$ and checks whether $\sigma$ is a valid for $m$.
For any valid pair of $gpk$ and $msk$ and any message $m$, correctness requires $\mathsf{TVerify}(gpk, \allowbreak m, \allowbreak \mathsf{TSign}(msk, m)) = 1$ ewnp. 
It also provides \emph{anonymity}: any two honestly generated signatures on the same message are indistinguishable, even if produced by different group members. 

$\mathsf{TOpen}$ and $\mathsf{TReveal}$ allow the GM to revoke the anonymity.
$\mathsf{TOpen}$ takes $gsk$ and $\sigma$ and returns the signer's identity. 
$\mathsf{TReveal}$ generates \emph{trapdoor} $td$.
Given $td$, $\mathsf{TTrace}$ allows users to determine whether any signature was issued by the same signer as $\sigma$.
\emph{Unforgeability} requires that no adversary can produce, ewnp., a valid signature on a message unsigned by a compromised signer, or a valid signature that cannot be opened or revealed to a compromised signer. 
\emph{Framing resistance} requires that no adversary can produce, ewnp., a valid signature that opens or reveals to an uncompromised signer.

\subsubsection{Undeniable Signature}
A 6-tuple of algorithms ($\mathsf{UGen}$, $\mathsf{USign}$, $\mathsf{UConfirm}$, $\mathsf{UDisavow}$, $\mathsf{UVerifyC}$, $\mathsf{UVerifyD}$) represents the non-interactive variant of the \emph{undeniable signature scheme}~\cite{chaum1989undeniable} in the random oracle model~\cite{jakobsson1996designated}.
The signer produces a signature $\sigma$ on a message $m$ by running $\mathsf{USign}(sk, m)$.
The validity of $\sigma$ cannot be determined without the signer's cooperation.
Instead, the signer generates a proof of the validity of $\sigma$ by $\upsilon \leftarrow \mathsf{UConfirm}(sk, m, \sigma)$ or the invalidity of $\sigma^\prime$ by $\bar{\upsilon} \leftarrow \mathsf{UDisavow}(sk, m, \sigma^\prime)$. 
These proofs are verified by checking $\mathsf{UVerifyC}(pk, m, \sigma, \upsilon) \ifeq 1$ and $\mathsf{UVerifyD}(pk, m, \sigma^\prime, \bar{\upsilon}) \ifeq 1$, respectively.
For every key pair $(sk, pk) \leftarrow \mathsf{UGen}(1^\kappa)$, message $m$, and $\sigma \leftarrow \mathsf{USign}(sk, m)$, correctness requires $\mathsf{UVerifyC}( pk, m, \sigma, \mathsf{UConfirm}(sk, m, \sigma)) = 1$ ewnp. 
For a random element $\sigma^\prime$ in the range of $\mathsf{USign}$, it holds ewnp. $\mathsf{UVerifyD}(pk, m, \sigma^\prime, \mathsf{UDisavow}(sk, m, \sigma^\prime)) = 1$. 

\emph{Unforgeability} requires that no adversary can produce a valid signature on an unsigned message, ewnp.
\emph{Undeniability} ensures that no signer can disavow a valid signature.
\emph{Invisibility} guarantees that the valid signature is indistinguishable from random elements in the signature space, without the signer's cooperation. 
Invisibility implies \emph{anonymity} of signers~\cite{galbraith2003invisibility}.

\subsubsection{One-way Authenticated Key Exchange}
A triplet of algorithms ($\mathsf{OGen}$, $\mathsf{OReply}$, $\mathsf{OAccept}$) represents the \emph{one-way authenticated key exchange protocol}~\cite{goldberg2013anonymity}. 
An \emph{unauthenticated peer} $A$ generates an ephemeral key pair $(x, X)$ by running $\mathsf{OGen}(1^\kappa)$. 
Similarly, an \emph{authenticated peer} $B$ generates $(y, Y)$ and a long-term key pair $(sk_B, pk_B)$.
They agree on the session key $k_{AB}$ by running $(k_{AB}, \sigma_B) \leftarrow \mathsf{OReply}(y, B, sk_B, X)$ at $B$ and then $k_{AB} \leftarrow \mathsf{OAccept}(x, B, pk_B, Y, \sigma_B)$ at $A$.

\emph{Session-key secrecy} ensures that the session key is held only by the authenticated peer and the unauthenticated peer who initiates the session, ewnp. 
\emph{One-way anonymity} means that the unauthenticated peer remains anonymous to the authenticated peer. \looseness=-1

\subsubsection{Key-Committing AE}
A pair of algorithms ($\mathsf{KEnc}$, $\mathsf{KDec}$) denotes the \emph{key-committing AE scheme}~\cite{farshim2017security,albertini2022abuse}, with the key generation algorithm left implicit.
For every key $k$ and message $m$, $\mathsf{KDec}(k, \mathsf{KEnc}(k, m)) = m$, ewnp. 
It ensures IND-CCA2 and \emph{key commitment}, which means that no adversary can find a ciphertext $c$ and two key-message pairs $(k, m) \neq (k^\prime, m^\prime)$ such that $m, m^\prime \neq \bot \land \mathsf{KDec}(k, c) = m \land \mathsf{KDec}(k^\prime, c) = m^\prime$, ewnp. 

%% file: rationale.tex
\section{Unforgeable and Non-repudiable Links}


\begin{figure}[t]
\centering
\includegraphics[width=1\linewidth]{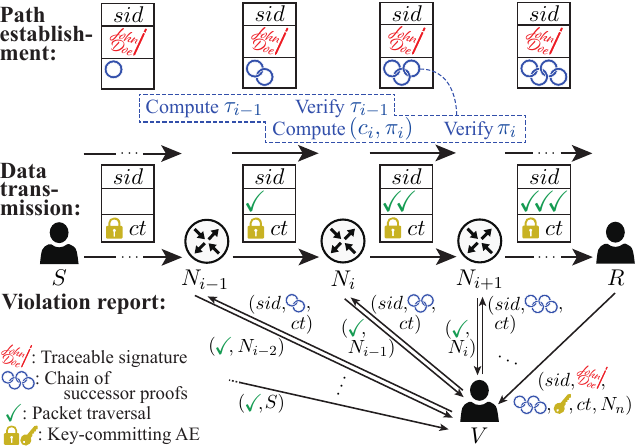}
\caption{Overview of the protocol and proof computation}
\vspace{-3mm}
\label{fig:2layer_sig}
\end{figure}

We refer to sender accountability as an \emph{unforgeable and non-repudiable link} from the message plaintext to the sender's identity, refining the definition in \cite{lee2016source}. 
However, establishing such a link \emph{directly} requires per-packet authorization~\cite{naylor2014balancing,lee2016source} or public-key cryptography~\cite{xu2012accountable,kopsell2006revocable,jonathancontractual,xia2020apgs}.
Instead, our approach establishes \emph{indirect} linkage from the plaintext to the sender's identity, using appropriate indirections to reduce expensive computation from per-packet to per-session. 
The indirections occur at three stages: \emph{sessions}, \emph{paths}, and \emph{packets}.

Our protocol comprises three phases: \emph{path establishment}, \emph{data transmission}, \emph{violation report}, each playing a distinct role in the linkage process, as shown in Figure~\ref{fig:2layer_sig}.
The sender establishes a connection to the receiver in the path establishment phase, and they exchange data in the data transmission phase.
If the receiver suspects a contract violation in the communication, it transitions to the violation report phase, where it submits the suspicious messages, along with the necessary transcripts obtained in the prior phases, to the verifier. 
If the request is legitimate, the verifier identifies the sender and provides the receiver with the sender's \emph{trapdoor}, which is used to determine whether any given traceable signature was produced by that sender. \looseness=-1

\subsection{Intuition for Linking Indirections}
\label{sec:intuition}
Indirection of the linkage from plaintext to sender's identity at the session level is obviously essential for reducing signatures from per-packet to per-session.
The reduced signatures establish the session-sender linkage, while the remaining linkage from plaintext to the session must be guaranteed by other means.

At the other end, packet-level indirection is also essential, since packets are encrypted. 
That is, the plaintext-packet linkage should naturally be ensured by the encryption scheme.

The remaining challenge is linking a transmitted packet to the session, which is nontrivial.
The protocol must verify that the packet was indeed transmitted, since a malicious receiver may craft a malicious plaintext, encrypt it into a packet, and report the packet to frame the sender.
Thus, a packet should not be directly linked to a session merely because it belongs to the session. 
Instead, an extra level of indirection at the path is needed: the verifier reconstructs the path used in the session and checks that the packet indeed traversed that path.

\subsection{Linking Sessions to Senders}
\label{sec:link1}
In the session establishment phase, the sender computes a traceable signature on a session identifier $sid$, which is the hash of her elliptic-curve Diffie-Hellman (ECDH) public key.
The receiver verifies this signature to ensure that the verifier, who plays the role of the GM, can open the sender's identity when needed; however, this verification does not reveal the sender's identity to the receiver, due to the anonymity of traceable signatures. \looseness=-1

\subsection{Linking Paths to Sessions}
\label{sec:link2}
We propose a cryptographic object, \emph{chain of successor proofs}.
This chain is computed by on-path relays in the path establishment phase, and suffices to convince the receiver that the verifier can reconstruct the path if all relays behave correctly, or otherwise a misbehaving relay is held accountable.

Such a chain is \emph{unforgeable}: an off-path adversary cannot produce a valid proof. 
It is also \emph{non-repudiable}: once a valid chain is created, on-path relays cannot deny their involvement without being blamed.
It is also \emph{secure against framing}: no honest relay can be falsely blamed for denial, and no honest sender can be framed as the sender on a
path it did not establish.
Finally, it is \emph{indistinguishable}: an honestly created chain leaks no knowledge of the path without the cooperation of the relays. 

As a foundation, we first introduce two types of proofs, a \emph{predecessor proof} and a \emph{successor proof}. 
Both are computed by two adjacent relays, say $N_{i-1}$ and $N_{i}$, to make their adjacency verifiable, using their key pairs $(sk_{i-1}, pk_{i-1})$ and $(sk_{i}, pk_{i})$. 
A predecessor proof is computed by $N_{i-1}$ to acknowledge itself as the predecessor of $N_{i}$ and that $N_{i+1}$ follows them in the session. 

\vspace{-1mm}
\begin{dfn}
\label{def:pred_proof}
A value $\tau_{i-1}$ is called a \emph{predecessor proof by} $N_{i-1}$ \emph{for} $N_{i}$ \emph{toward} $N_{i+1}$ \emph{in} session $sid$:
\vspace{-0.3mm}
\setlength{\leftmargini}{0.5em}
\begin{itemize}
    \item[] \textbf{Computation}: $\tau_{i-1} \leftarrow \mathsf{Sign}(sk_{i-1}, sid \parallel N_{i} \parallel N_{i+1})$.
    \item[] \textbf{Verification}: Check $\mathsf{Verify}(pk_{i-1}, sid \parallel N_{i} \parallel N_{i+1}, \tau_{i-1}) \ifeq 1$. 
\end{itemize}
\end{dfn}

\vspace{-3mm}
\begin{prop}
\label{thm:pred_proof}
A predecessor proof is \emph{unforgeable} and \emph{non-repudiable}.
\vspace{-1.5mm}
\end{prop}
\begin{proofsketch}
Unforgeability is proven by reduction to the unforgeability of the signature scheme. 
Non-repudiation follows from the fact that the signature is universally verifiable.
\vspace{-1.0mm}
\end{proofsketch}

A successor proof is computed by $N_{i}$, given $\tau_{i-1}$ from $N_{i-1}$, to acknowledge their adjacency along the path and that $N_{i+1}$ follows them in the session. 
It is also bound to an auxiliary input. 
Including the identity of the two-hop-ahead node in the proof input prevents a local adversary from diverting the verifier’s tracing from a maliciously established path to an honestly established path. \looseness=-1

\vspace{-1mm}
\begin{dfn}
\label{def:succ_proof}
A pair $(c_{i}, \pi_{i})$ is called a \emph{successor proof by} $N_{i}$ \emph{for} $N_{i-1}$ \emph{toward} $N_{i+1}$ \emph{with} auxiliary input ${aux}_{i}$ \emph{in} session $sid$:
\vspace{-0.3mm}
\setlength{\leftmargini}{0.5em}
\begin{itemize}
    \item[] \textbf{Computation}: $N_{i}$ verifies $\tau_{i-1}$, samples $r_{i}$ from $\{0,1\}^\kappa$, and invokes $c_{i} \leftarrow \mathsf{Com}(r_{i}, \tau_{i-1})$; $\pi_{i} \leftarrow \mathsf{USign}(sk_{i}, {aux}_{i} \parallel c_{i})$.
    \item[] \textbf{Verification}: The verifier requests $(\upsilon_i, r_i, \tau_{i-1}, N_{i-1})$ from $N_i$, such that the confirmation $\upsilon_{i} = \mathsf{UConfirm}(sk_i, {aux}_{i} \parallel c_{i}, \pi_i)$, and checks $\mathsf{UVerifyC}(pk_{i}, {aux}_{i} \parallel c_{i}, \pi_{i}, \upsilon_{i}) \ifeq 1$, $\mathsf{Com}(r_{i}, \tau_{i-1}) \ifeq c_{i}$, and the verification of $\tau_{i-1}$ succeeds. For any request involving an invalid $(c_{i}^\prime, \pi_{i}^\prime)$ and ${aux}^\prime$, $N_{i}$ instead returns a disavowal $\bar{\upsilon}_{i} \gets \mathsf{UDisavow}(sk_i, {aux}^\prime  \parallel c_i^\prime,\pi_i^\prime)$, which is verified by checking $\mathsf{UVerifyD}(pk_{i}, {aux}_{i}^\prime \parallel c_{i}^\prime, \pi_{i}^\prime, \bar{\upsilon}_{i}) \ifeq 1$.
\end{itemize}
\end{dfn}
\vspace{-3mm}
\begin{prop}
\label{thm:succ_proof}
A successor proof is \emph{unforgeable}, \emph{undeniable}, and \emph{indistinguishable}. 
Indistinguishability here requires that, for any pair of secret keys $sk_i^0, sk_i^1$, any pair of predecessor proofs $\tau_{i-1}^0, \tau_{i-1}^1$, and any common auxiliary input ${aux}_i$, the honestly computed successor proofs, $(c_i^0, \pi_i^0, {aux}_i)$ and $(c_i^1, \pi_i^1, {aux}_i)$, are computationally indistinguishable. \
\vspace{-1.0mm}
\end{prop}
\begin{proofsketch}
Undeniability follows from that of the underlying undeniable signature scheme.
Unforgeability follows from that of the undeniable signature scheme and the binding property of the commitment scheme.
Indistinguishability follows from the invisibility of the undeniable signature scheme and the hiding property of the commitment scheme.
\vspace{-1.0mm}
\end{proofsketch}

We next define a \emph{chain of successor proofs}, computed by all on-path relays.
Its verification enables reconstruction of the path in reverse order, from the receiver to the sender. 

\vspace{-1mm}
\begin{dfn}
\label{def:succ_proof_chain}
$(C, \Pi) := ((c_1, ..., c_{n}), (\pi_0, ..., \pi_{n}))$ is called a \emph{chain of successor proofs} \emph{for} path $N_0, ..., N_{n+1}$ in session $sid$. 
\vspace{-0.5mm}
\setlength{\leftmargini}{0.5em}
\begin{itemize}
    \item[] \textbf{Computation}: $N_0$ computes $\pi_0 \gets \mathsf{USign}(sk_0, aux_0)$ with $aux_0 = sid$. For each $i$ from $1$ to $n$, $N_{i}$ requests from $N_{i-1}$ a confirmation $\rho_{i-1} = \mathsf{UConfirm}(sk_i, {aux}_{i} \parallel c_{i}, \pi_i)$ and $\tau_{i-1}$, checks $\mathsf{UVerifyC}(pk_{i-1}, aux_{i-1} \parallel c_{i-1}, \pi_{i-1}, \rho_{i-1}) \ifeq 1$, and computes $(c_{i}, \pi_{i})$ for $N_{i-1}$ toward $N_{i+1}$ with ${aux}_{i} := sid \parallel \pi_0 \parallel ... \parallel \pi_{i-1} \parallel c_1 \parallel ... \parallel c_{i-1}$.
    \item[] \textbf{Verification}: The verifier queries each $N_{i}$ to verify the corresponding $(c_{i}, \pi_{i})$, for each $i$ from $n$ downto $1$. 
    $N_i$ returns either $(\upsilon_{i}$, $r_{i}$, $\tau_{i-1}$, $N_{i-1})$ for confirmation or $\bar{\upsilon}_{i}$ for disavowal. If the former is valid, the verifier then proceeds to query $N_{i-1}$. If the latter is valid, the verifier identifies $N_{i+1}$ as a malicious relay. Otherwise, $N_{i}$ is identified as a malicious relay. 
\end{itemize}
\end{dfn}
\vspace{-3mm}
\begin{thm}
\label{thm:succ_proof_chain}
A chain of successor proofs is \emph{unforgeable} and \emph{indistinguishable}. 
\vspace{-1.5mm}
\end{thm}
\begin{proofsketch}
Unforgeability and indistinguishability follow from these properties of the constituent successor proofs. 
\vspace{-1.0mm}
\end{proofsketch}

We see that non-repudiation is achieved by requiring relays to either confirm or disavow a successor proof, and the trusted verifier to identify a malicious relay based on their responses.
We also see that no honest relay can be falsely blamed as malicious.
\vspace{-4.5mm}
\begin{thm}
\label{thm:chain_nonrep}
Given that the verifier is trusted, a chain of successor proofs is \emph{non-repudiable}. 
\vspace{-1.5mm}
\end{thm}
\begin{proofsketch}
Non-repudiation follows from the undeniability of the undeniable signature scheme and the fact that any relay that submits neither a valid confirmation nor a valid disavowal is blamed as malicious. 
No relay can evade responsibility by falsifying the chain: any alteration is detected by the nearest upstream honest relay, which can then submit a valid disavowal.
This allows the verifier to identify the last malicious relay. 
\vspace{-1.0mm}
\end{proofsketch}

\vspace{-2mm}
\begin{thm}
\label{thm:chain_framing}
Given that the verifier is trusted, an honest relay is not blamed as a malicious relay, ewnp.
\vspace{-1.5mm}
\end{thm}
\begin{proofsketch}
An honest relay can submit either a valid confirmation or disavowal, regardless of how other relays behave. \looseness=-1
\vspace{-1.0mm}
\end{proofsketch}

A chain of successor proofs does not guarantee that $N_{i-1}$ and $N_{i}$ are indeed on the path. 
Two relays---either malicious or chosen by a malicious sender---may falsely claim to belong to an honest path by diverting the chain verification process on that honest path onto another malicious path they actually belong to, collaboratively crafting a successor proof that is valid yet bridges these distinct paths, as detailed in Section~\ref{sec:misdirection}.
Nevertheless, an honest sender remains protected from framing. \looseness=-1

\vspace{-1mm}
\begin{thm}
\label{thm:chain_branch}
Under the local adversary assumption on an honestly established path, an honest sender cannot be held responsible for a path that it did not actually establish, ewnp.
\vspace{-1.0mm}
\end{thm}
\begin{proofsketch}
We argue by contradiction. 
Consider an honest sender $S$ establishes a path $(S, N_1, ..., N_{n}, R)$, while a diverted path $(S, N_1^\prime,...,N_n^\prime, R^\prime)$, not actually established by $S$, is reconstructed by the verifier. 
Let $i \geq 2$ be the first position at which $N_i \neq N_i^\prime$. 
For the path reconstruction along the diverted path from $N_i^\prime$ to reach $S$, both $N_{i-1}$ and $N_{i-2}$ must acknowledge that $N_i^\prime$, rather than $N_i$, follows them.
The two malicious relays on the honestly established path contradict the assumption. 
\vspace{-1.0mm}
\end{proofsketch}

\subsection{Linking Packets to Paths}
\label{sec:link3}
In the violation report phase, the verifier queries the relays to determine whether the packet traversed the path. 
The verifier must not be misled by dishonest responses or malicious relays that alter the queried packet, which cause even honest relays to return invalid responses.
To this end, relays ensure the integrity of forwarded packets in the data transmission phase, like \emph{path validation}~\cite{naous2011verifying, kim2014lightweight}, and the verifier consolidates responses via \emph{majority voting} under the honest majority assumption. \looseness=-1

In the data transmission phase, for every packet and each relay $N_i$, two MACs are computed over the conveyed ciphertext $ct$: $m_{Si}$, computed by the sender and verified by $N_i$ using the shared key $k_{Si}$, and $m_{iR}$, computed by $N_i$ and verified by the receiver using $k_{iR}$. 
Thus, the relay can ensure that the sender transmitted $ct$ and that the receiver will accept it only if it remains unaltered.
The former prevents a malicious receiver and colluding relays from framing a sender, while the latter prevents a malicious sender and colluding relays from injecting malicious packets.

\vspace{-1mm}
\begin{prop}
\label{thm:data_mac}
Assuming that the relay $N_i$ is honest, and the session-key secrecy of $k_{Si}$ and $k_{iR}$ holds, the data transmission phase guarantees the following properties, ewnp:
\setlength{\leftmargini}{1.2em}
\begin{itemize}
    \item If the sender is honest, $N_i$ only forwards the ciphertext that is transmitted by the sender.
    \item If the receiver is honest, the receiver accepts only the ciphertext that is forwarded by $N_i$.
\end{itemize}
\end{prop}

$k_{Si}$ and $k_{iR}$ are securely established in the path establishment phase. 
To establish $k_{Si}$, the sender and $N_i$ perform an ECDH key exchange in which the sender authenticates $N_i$ using its certificate.
To establish $k_{iR}$, the list $K$ of all relays' ECDH public keys is conveyed to the receiver, and the relays include $K$ in the auxiliary input of the chain of successor proofs $(C, \Pi)$.
This allows the verifier in the violation report phase to detect any alteration of $K$, thereby preventing man-in-the-middle attacks.

\vspace{-1mm}
\begin{prop}
\label{thm:data_key}
The path establishment phase guarantees the session-key secrecy of $k_{Si}$ between the sender and the relay $N_i$, and $k_{iR}$ between $N_i$ and the receiver, ewnp.
\end{prop}

In the violation report phase, the verifier queries each $N_{i}$ with the message ciphertext $ct$.
Upon receiving the query, $N_{i}$ reports whether it had forwarded $ct$ in the data transmission phase. 
The verifier collects responses from all $n$ on-path relays, and if a majority of them respond affirmatively, it concludes that the packet has indeed traversed the path. 

\vspace{-1mm}
\begin{thm}
\label{thm:data_all}
Under the honest majority assumption, the path establishment, data transmission, and violation report phases guarantee that the packet–path links can neither be forged against an honest sender nor repudiated against an honest receiver, ewnp.
\vspace{-1.5mm}
\end{thm}
\vspace{-5mm}
\begin{proofsketch}
By Propositions~\ref{thm:data_mac} and \ref{thm:data_key}, if the sender is honest, a majority of relays respond affirmatively only for ciphertexts transmitted by the sender. If the receiver is honest, it accepts only ciphertexts for which a majority responds affirmatively.
Thus, ciphertexts affirmatively concluded by the verifier are a subset of those transmitted by an honest sender, and are a superset of those accepted by an honest receiver.
\end{proofsketch}

\subsection{Linking Plaintexts to Packets}
\label{sec:link4}
The sender in the data transmission phase encrypts the plaintext $pt$ with the session key $k_{SR}$ using a key-committing AE scheme, and sends the ciphertext $ct$ to the receiver. 
If $pt$ is suspected in the violation report phase, the receiver submits $pt$ together with $ct$ and $k_{SR}$ to the verifier, who verifies that $ct$ is correctly decrypted into $pt$.
The key-committing property ensures that the verifier can uniquely link the plaintext to the ciphertext, even if a malicious receiver arbitrarily selects the key. \looseness=-1

%% file: protocol.tex
\section{\pname: Proposed Protocol}
\label{sec:protocol}

\begin{table}[t]
\fontsize{8pt}{8pt}\selectfont
\caption{Notation used in protocol description}
\renewcommand{\arraystretch}{1.15}
\label{tab:notation}
    \centering
    \begin{tabular}{>{\centering\arraybackslash}p{12mm}p{60mm}}
    \toprule
    Symbol  & Description \\
    \midrule
    $\kappa$            & Security parameter \\
    $G, q$              & Base point and its order in the elliptic curve group \\
    $sk_i, pk_i$        & Long-term secret and public keys of $N_i$ \\
    $msk_S, td_S$       & Long-term member secret key and trapdoor of $S$ \\
    $gsk, gpk$          & Long-term group secret and public keys \\
    $x_i$               & Ephemeral ECDH public keys of $N_i$ \\
    $k_{Si}$            & Session Key between $S$ and $N_i$ \\
    $k_{iR}$            & Session Key between $N_i$ and $R$ \\
    $k_{SR}$            & Session key between $S$ and $R$ for data encryption \\
    $\sigma_S, \sigma_R$ & Traceable signature of $S$ and OWAKE signature of $R$ \\
    $\mathscr{C}_R$     & Contract specified and published by $R$ \\
    $ts$                & Timestamp at the path establishment \\
    $sid$          & Session identifier \\
    $e_i$          & Encrypted forwarding information for $N_i$ \\
    $\tau_i$       & Predecessor proof by $N_i$ for $N_{i+1}$ toward $N_{i+2}$ \\ 
    $(c_i, \pi_i)$ & Successor proof by $N_i$ for $N_{i-1}$ toward $N_{i+1}$ \\
    $r_i$          & Randomness used by $N_i$ in calculating $c_i$ \\ 
    $\rho_i$       & Confirmation of $\pi_i$ by $N_i$ for $N_{i+1}$ \\
    $\upsilon_i, \bar{\upsilon}_i$  & Confirmation and disavowal of $\pi_i$ by $N_i$ for $V$ \\
    $(C, \Pi)$     & Chain of successor proofs \\
    $E, K, M$      & Lists of $e_i, x_iG, m_{Si}$ and $m_{iR}$ in packet headers \\
    $pt, ct$       & Plaintext and ciphertext of message \\
    $m_{Si}, m_{iR}$   & MACs of ciphertext calculated by $N_i$ \\
    $h, p$         & Packet header and payload \\
    $\mathcal{S}_i$ & State maintained by $N_i$ \\
    $\epsilon, {} \parallel {}$ & Empty string and string concatenation \\
    \bottomrule
    \end{tabular}
    \vspace{0mm}
\end{table}

\begin{figure}[t]
\centering
    \begin{minipage}[b]{0.95\linewidth}
    \adjustimage{scale=0.38,left}{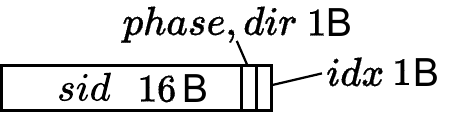}
    \subcaption{Common header}
    \label{fig:header1}
    \end{minipage} \\ \vskip3mm
    \begin{minipage}[b]{0.95\linewidth}
    \hspace{-0.1mm}\adjustimage{scale=0.38,left}{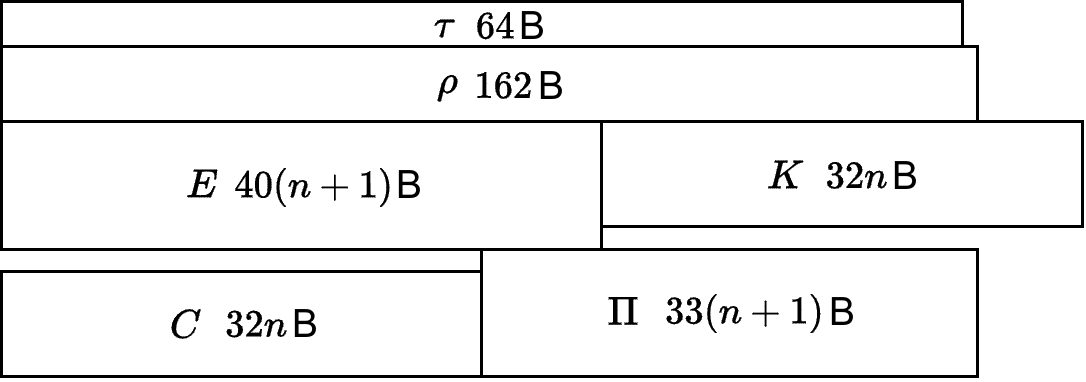}
    \subcaption{Header (path establishment phase)}
    \label{fig:header2}
    \end{minipage} \\  \vskip3mm
    \begin{minipage}[b]{0.95\linewidth}
    \adjustimage{scale=0.38,left}{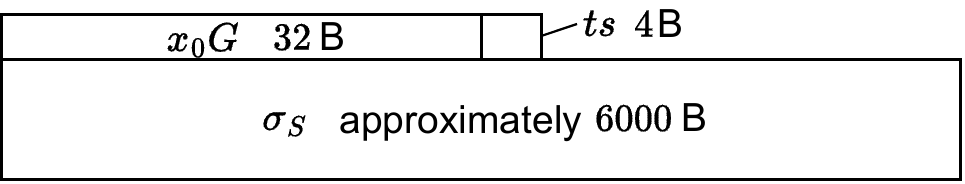}
    \subcaption{Payload (path establishment phase, forward)}
    \label{fig:header3}
    \end{minipage} \\  \vskip3mm
    \begin{minipage}[b]{0.95\linewidth}
    \adjustimage{scale=0.38,left}{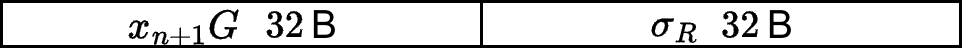}
    \subcaption{Payload (path establishment phase, backward)}
    \label{fig:header4}
    \end{minipage} \\  \vskip3mm
    \begin{minipage}[b]{0.95\linewidth}
    \adjustimage{scale=0.38,left}{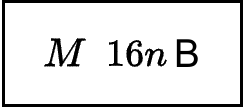}
    \subcaption{Header (data transmission phase, forward)}
    \label{fig:header5}
    \end{minipage}
\vspace{1mm}
\caption{The \pname packet format}
\vspace{1mm}
\label{fig:header}
\end{figure}

In the following, the header fields are indicated with rectangles (e.g., \smash{$\fdsid$}\,). \smash{$\fdsrc$} and \smash{$\fddst$} denote the source and destination address fields in the lower-layer (e.g., IP) header. 
Table~\ref{tab:notation} summarizes the notation used in the protocol description throughout the paper.

\subsection{Packet Format}
Figure~\ref{fig:header} illustrates the packet format of \pname in the path establishment and data transmission phases.
Packets in both phases share a common header, shown in Figure~\ref{fig:header1}, at the beginning of the encrypted TLS payload. 
The fields following this header differ depending on the phase and direction.
In the path establishment phase, the common header is followed by the header shown in Figure~\ref{fig:header2} and the payload shown in either Figure~\ref{fig:header3} or Figure~\ref{fig:header4}, depending on the direction.
When $\sigma_S$ exceeds the Maximum Transmission Unit (MTU), it is split across multiple packet payloads. 
Instead, short traceable signatures based on pairings~\cite{nguyen2004efficient,choi2006short} may be used in place of the RSA-based construction. 

In the data transmission phase, the common header is followed by the header shown in Figure~\ref{fig:header5} only in the forward direction.
The payload carries the encrypted data.

\subsection{Setup and Registration}
Upon the system's boot-up, the verifier $V$ runs initialization once as a GM: $(gsk, gpk) \leftarrow \mathsf{TSetup}(1^\kappa)$ and publishes $gpk$. 

To join the system, a new party obtains $msk\leftarrow \mathsf{TJoin}(1^\kappa)$ through an interaction with the GM.
It also generates key pairs by running $\mathsf{Gen}(1^\kappa)$, $\mathsf{UGen}(1^\kappa)$, and $\mathsf{OGen}(1^\kappa)$.
For simplicity, we collectively denote these key pairs as $(sk, pk)$, while using the appropriate pair for each algorithm in practice.
If the party intends to act as a receiver, it also specifies a contract.
Finally, it publishes the resulting public keys and contract in the public directory.
The party can rotate its keys and contract at any time. \looseness=-1

  \begin{algorithm}[t]
    \scriptsize
    \captionsetup{font=small}
    \caption{~Sender $S$ in the path setup phase}
    \label{alg:path_sender}
    \begin{algorithmic}
\Procedure{pathForward}{$N_1, ..., N_n, R$}
    \State \Assert $n \geq 3 \land \left| \{S, N_1, ..., N_n, R\}\right| = n+2$
    \State $x_0 \sample [1, ..., q - 1]$; \quad $sid \gets \mathsf{H}(x_0G)$
    \State \Assert $\forall ({sid}^\prime, \_) \in \mathcal{S}_i$:\; ${sid}^\prime \neq sid$
    \State $ts \gets \mathsf{now}()$; \quad $\sigma_S \gets \mathsf{TSign}(msk_S, x_0G \parallel ts)$
    \For{$i = 1$ to $n$}
        \State $k_{Si} \gets x_0 \times pk_i$; \hspace{0.5em} $e_i \gets \mathsf{Enc}(k_{Si}, n \parallel i \parallel N_{i-1} \parallel N_{i + 1} \parallel N_{i+2})$ \Comment{$N_{n+2} := \bot$}
    \EndFor
    \State $k_{Sn+1} \gets x_0 \times pk_{n+1}$; \quad $e_{n + 1} \gets \mathsf{Enc}(k_{Sn+1}, n \parallel (n+1) \parallel N_{n})$
    
    \State Initialize $h$; \quad $\fdsrc \gets S$; \quad $\fddst \gets N_1$; \quad $\fdsid \gets sid$
    \State $\fdidx \gets 0$; \quad $\fdphase \gets \mathtt{Path}$; \quad $\fddir \gets \mathtt{Forward}$
    \State $\fdE \gets e_1 \parallel \cdots \parallel e_{n+1}$; \quad $\fdK \gets \epsilon$; \quad $\fdC \gets \epsilon$

    \State $\pi_0 \gets \mathsf{USign}(sk_0, sid)$; \quad $\fdPi \gets \pi_0$
    \State $\rho_0 \gets \mathsf{UConfirm}(sk_0, sid, \pi_0)$; \quad $\fdrho \gets \rho_0$
    
    \State $\tau_0 \gets \mathsf{Sign}(sk_0, sid \parallel N_1 \parallel N_2)$; \quad $\fdtau \gets \tau_0$
    \State $p \gets x_0G \parallel ts \parallel \sigma_S$

    \If{no TLS connection exists between $S$ and $N_1$}
        \State Invoke the TLS handshake protocol with $N_1$
    \EndIf
    \State \begin{varwidth}[t]{\linewidth}$
    \mathcal{S}_0 \gets \mathcal{S}_0 \cup \{(sid, \mathtt{Pending}, n, 0, N_1, ..., N_n, R, ts, $ \par
    \hskip17mm $k_{S0}, ..., k_{Sn}, x_0)\}$
    \end{varwidth}
    \State Transmit $(h \parallel p)$ to $N_1$ using the TLS record protocol
\EndProcedure
\vspace{1mm}
\Procedure{pathBackward}{$h \parallel p$}
    \State \Assert $\fdphase = \mathtt{Path} \land \fddir = \mathtt{Backward} \land \fdidx = 0$
    \State $N_1 \gets \fdsrc$\,; \quad $sid \gets \fdsid$\,; \quad $x_{n+1}G \parallel \sigma_R \gets p$ 
    \State \begin{varwidth}[t]{\linewidth}
        \Assert $\exists n, N_2, ..., N_n, R, ts, k_{S0}, ..., k_{Sn}, x_0$: \par
        \hskip3mm $(sid, \mathtt{Pending}, n, 0, N_1, ..., N_n, R, ts, k_{S0}, ..., k_{Sn}, x_0) \in \mathcal{S}_0$
        \end{varwidth}
    \State $k_{SR} \gets \mathsf{OAccept}(x_0, R, pk_{n+1}, x_{n+1}G, \sigma_R)$
    \State \begin{varwidth}[t]{\linewidth}
        $\mathcal{S}_0 \gets \mathcal{S}_0 \setminus \{(sid, \mathtt{Pending}, n, 0, N_1, ..., N_n, R, ts, k_{S0}, ..., k_{Sn}, x_0)\}$ \par
        \hskip9mm ${} \cup \{(sid, \mathtt{OK}, n, 0, N_1, ..., N_n, R, ts, k_{S0}, ..., k_{Sn}, k_{SR})\}$
        \end{varwidth}
    \State Deliver $(sid, N_1, ..., N_n, R)$ to the upper layer
\EndProcedure
    \end{algorithmic}
  \end{algorithm}

  \begin{algorithm}[t]
    \scriptsize
    \captionsetup{font=small}
    \caption{~Relay $N_i$ in the path setup phase}
    \label{alg:path_relay}
    \begin{algorithmic}
\Procedure{pathForward}{$h \parallel p$}
    \State \Assert $\fdphase = \mathtt{Path} \land \fddir = \mathtt{Forward} \land \#\fdE > 1$
    \State $sid \gets \fdsid$\,; \quad $x_0G \parallel \_ \gets p$
    \State \Assert $sid = \mathsf{H}(x_0G) \land \forall ({sid}^\prime, \_) \in \mathcal{S}_i$:\; ${sid}^\prime \neq sid$
    \State $k_{Si} \gets sk_i \times x_0G$; \quad $e_i \parallel \fdE \gets \fdE$
    \State $n \parallel i \parallel N_{i - 1} \parallel N_{i + 1} \parallel N_{i+2} \gets \mathsf{Dec}(k_{Si}, e_i)$
    \State \Assert $1 \leq i \leq n \land \fdsrc = N_{i - 1}$
    \State \Assert $\fdidx = \#\fdK = \#\fdC = \#\fdPi = n + 1 - \#\fdE = i-1$
    \State $\tau_{i - 1} \gets \fdtau$\,; \quad $\hat{\Pi} \parallel \pi_{i-1} \gets \fdPi$\,; \quad $\rho_{i - 1} \gets \fdrho$
    \State \Assert $\mathsf{Verify}(pk_{i - 1}, sid \parallel N_i \parallel N_{i+1}, \tau_{i - 1}) = 1$
    \State \Assert $\mathsf{UVerifyC}(pk_{i-1}, sid \parallel \fdK \parallel \fdC \parallel \hat{\Pi}, \pi_{i - 1}, \rho_{i - 1}) \hspace{0.5ex}{=}\hspace{0.5ex} 1$
    \State $x_i \sample [1, ..., q - 1]$; \quad $\fdK \gets \fdK \parallel x_iG$
    \State $r_i \sample \{0, 1\}^\kappa$; \quad $c_i \gets \mathsf{Com}(r_i, \tau_{i-1})$; \quad $\fdC \gets \fdC \parallel c_i$
    \State $\pi_{i} \gets \mathsf{USign}(sk_{i}, sid \parallel \fdK \parallel \fdC \parallel \fdPi)$
    \State $\rho_i \gets \mathsf{UConfirm}(sk_i, sid \parallel \fdK \parallel \fdC \parallel \fdPi\,, \pi_{i})$; \quad $\fdrho \gets \rho_{i}$
    \State $\fdPi \gets \fdPi \parallel \pi_i$; \quad $\tau_i \gets \mathsf{Sign}(sk_i, sid \parallel N_{i + 1} \parallel N_{i+2})$; \quad $\fdtau \gets \tau_i$
    \State $\fdsrc \gets N_i$; \quad $\fddst \gets N_{i + 1}$; \quad $\fdidx \gets i$
    \If{no TLS connection exists between $N_i$ and $N_{i + 1}$}
        \State Invoke the TLS handshake protocol with $N_{i + 1}$
    \EndIf
    \State \begin{varwidth}[t]{\linewidth}
    $\mathcal{S}_i \gets \mathcal{S}_i \cup \{(sid, \mathtt{Pending}, n, i, N_{i - 1}, N_{i + 1}, N_{i+2}, k_{Si}, x_i, \tau_{i - 1}, r_i, \varnothing)\}$
    \end{varwidth}
    \State Transmit $(h \parallel p)$ to $N_{i + 1}$ using the TLS record protocol
\EndProcedure
\vspace{1mm}
\Procedure{pathBackward}{$h \parallel p$}
    \State \Assert $\fdphase = \mathtt{Path} \land \fddir = \mathtt{Backward} \land \fdidx > 0$
    \State $N_{i+1} \gets \fdsrc$\,; \hskip0.5em $sid \gets \fdsid$\,; \hskip0.5em $i \gets \fdidx$\,; \hskip0.5em $x_{n+1}G \parallel \_ \gets p$
    \State \begin{varwidth}[t]{\linewidth}
        \Assert $\exists n, N_{i-1}, N_{i+2}, k_{Si}, x_i, \tau_{i-1}, r_i$: \par
        \hskip10mm $(sid, \mathtt{Pending}, n, i, N_{i - 1}, N_{i + 1}, N_{i+2}, k_{Si}, x_i, \tau_{i-1}, r_i, \varnothing) \in \mathcal{S}_i$
        \end{varwidth}

    \State $k_{iR} \gets x_i \times x_{n+1}G$
    
    \State \begin{varwidth}[t]{\linewidth}
        $\mathcal{S}_i \gets \mathcal{S}_i \setminus \{(sid, \mathtt{Pending}, n, i, N_{i - 1}, N_{i + 1}, N_{i+2}, k_{Si}, x_i, \tau_{i-1}, r_i, \varnothing)\}$ \par
        \hskip6mm ${} \cup \{(sid, \mathtt{OK}, n, i, N_{i - 1}, N_{i + 1}, N_{i+2}, k_{Si}, k_{iR}, \tau_{i-1}, r_i, \varnothing)\}$
        \end{varwidth}
    \State $\fdsrc \gets N_i$; \quad $\fddst \gets N_{i - 1}$; \quad $\fdidx \gets i - 1$
    \State Transmit $(h \parallel p)$ to $N_{i - 1}$ using the TLS record protocol
\EndProcedure
    \end{algorithmic}
  \end{algorithm}

  \begin{algorithm}[t]
    \scriptsize
    \captionsetup{font=small}
    \caption{~Receiver $R$ in the path setup phase}
    \label{alg:path_receiver}
    \begin{algorithmic}
\Procedure{pathReply}{$h \parallel p$}
    \State \Assert $\fdphase = \mathtt{Path} \land \fddir = \mathtt{Forward} \land \#\fdE = 1$
    \State $sid \gets \fdsid$\,; \quad $x_0G \parallel ts \parallel \sigma_S \gets p$
    \State \Assert $ts$ is fresh
    \State \Assert $sid = \mathsf{H}(x_0G) \land \forall ({sid}^\prime, \_) \in \mathcal{S}_i$:\; ${sid}^\prime \neq sid$
    \State $k_{Sn+1} \gets sk_{n+1} \times x_0G$
    \State \Assert $\mathsf{TVerify}(gpk, x_0G \parallel ts, \sigma_S)$

    \State \Assert $\forall (\_, \mathtt{Trapdoor}, td) \in \mathcal{S}_{n+1}$:\; $\mathsf{TTrace}(\sigma_S, td) = 0$
    
    \State $e_{n + 1} \gets \fdE$\,; \quad $n \parallel (n + 1) \parallel N_n \gets \mathsf{Dec}(k_{Sn+1}, e_{n + 1})$
    \State \Assert $\fdsrc = N_n \land \fdidx = \#\fdK = \#\fdC = \#\fdPi -1 = n$
    \State $\tau_{n} \gets \fdtau$\,; \quad $\rho_{n} \gets \fdrho$
    \State $K \gets \fdK$\,; \quad $C \gets \fdC$\,; \quad $\Pi \gets \fdPi$\,; \quad $\hat{\Pi} \parallel \pi_{n} \gets \Pi$
    \State \Assert $\mathsf{Verify}(pk_{n}, sid \parallel R, \tau_{n}) = 1$
    \State \Assert $\mathsf{UVerifyC}(pk_{n}, sid \parallel K \parallel C \parallel \hat{\Pi}, \pi_{n}, \rho_{n}) = 1$

    \State $x_{n+1} \sample [1, ..., q - 1]$; \quad $k_{SR}, \sigma_R \gets \mathsf{OReply}(x_{n+1}, R, sk_{n+1}, x_0G)$

    \State $x_1G \parallel ... \parallel x_nG \gets K$
    \For{$i = n$ \textbf{downto} $1$}
        \State $k_{iR} \gets x_{n + 1} \times x_iG$
    \EndFor
    
    \State Initialize $h$; \quad $\fdsrc \gets R$; \quad $\fddst \gets N_n$; \quad $\fdsid \gets sid$
    \State $\fdidx \gets n$; \quad $\fdphase \gets \mathtt{Path}$; \quad $\fddir \gets \mathtt{Backward}$
    \State $p \gets x_{n+1}G \parallel \sigma_R$
    
    \State \begin{varwidth}[t]{\linewidth}
        $\mathcal{S}_{n + 1} \gets \mathcal{S}_{n + 1} \cup \{(sid, \mathtt{OK}, n, n + 1, N_{n}, ts, k_{SR}, k_{1R}, ..., k_{nR}, $  \par
        \hskip19mm $\tau_n, x_0G, \sigma_S, K, C, \Pi)\}$
        \end{varwidth}
        
    \State Transmit $(h \parallel p)$ to $N_n$ using the TLS record protocol
    \State Deliver $sid$ to the upper layer
\EndProcedure
    \end{algorithmic}
  \end{algorithm}

  \begin{algorithm}[t]
    \scriptsize
    \captionsetup{font=small}
    \caption{~Sender $S$ in the data transmission phase}
    \label{alg:data_sender}
    \begin{algorithmic}
\Procedure{dataForward}{$sid, pt$}
    \State \begin{varwidth}[t]{\linewidth}
        \Assert $\exists n, N_1, ..., N_n, R, ts, k_{SR}$: \par
        \hskip8mm $(sid, \mathtt{OK}, n, 0, N_1, ..., N_n, R, ts, k_{S0}, ..., k_{Sn}, k_{SR}) \in \mathcal{S}_0$
        \end{varwidth}
    \State \Assert $\mathscr{C}_R(ts, pt) = 1$; \quad $ct \gets \mathsf{KEnc}(k_{SR}, pt)$
    \State Initialize $h$; \quad $\fdsrc \gets S$; \quad $\fddst \gets N_1$; \quad $\fdsid \gets sid$
    \State $\fdidx \gets 0$; \quad $\fdphase \gets \mathtt{Data}$; \quad $\fddir \gets \mathtt{Forward}$
    \For{$i = 1$ \textbf{to} $n$}
        \State $m_{Si} \gets \mathsf{MAC}(k_{Si}, ct)$
    \EndFor
    \State $\fdM \gets m_{S1} \parallel ... \parallel m_{Sn}$
    \State Transmit $(h \parallel ct)$ to $N_1$ using the TLS record protocol
\EndProcedure
\vspace{1mm}
\Procedure{dataBackward}{$h \parallel p$}
    \State \Assert $\fdphase = \mathtt{Data} \land \fddir = \mathtt{Backward} \land \fdidx = 0$
    \State $N_1 \gets \fdsrc$\,; \quad $sid \gets \fdsid$\,; \quad $ct \gets p$
    \State \begin{varwidth}[t]{\linewidth}
        \Assert $\exists n, N_2, ..., N_n, R, ts, k_{SR}$: \par
        \hskip8mm $(sid, \mathtt{OK}, n, 0, N_1, ..., N_n, R, ts, k_{S0}, ..., k_{Sn}, k_{SR}) \in \mathcal{S}_0$
        \end{varwidth}
    \State $pt \gets \mathsf{Dec}(k_{SR}, ct)$; \quad Deliver $(sid, pt)$ to the upper layer
\EndProcedure
    \end{algorithmic}
  \end{algorithm}

  \begin{algorithm}[t]
    \scriptsize
    \captionsetup{font=small}
    \caption{~Relay $N_i$ in the data transmission phase}
    \label{alg:data_relay}
    \begin{algorithmic}
\Procedure{dataForward}{$h \parallel p$}
    \State \Assert $\fdphase = \mathtt{Data} \land \fddir = \mathtt{Forward}$
    \State $N_{i - 1} \gets \fdsrc$\,; \quad $sid \leftarrow \fdsid$\,; \quad $i \gets \fdidx + 1$; \quad $ct \gets p$
    \State \begin{varwidth}[t]{\linewidth}
        \Assert $\exists n, N_{i+1}, N_{i+2}, k_{Si}, k_{iR}, \tau_{i-1}, r_i, \mathcal{P}$: \par
        \hskip5mm $n \geq i \land (sid, \mathtt{OK}, n, i, N_{i - 1}, N_{i + 1}, N_{i+2}, k_{Si}, k_{iR}, \tau_{i-1}, r_i, \mathcal{P}) \in \mathcal{S}_i$
        \end{varwidth}

    \State \Assert $\#\fdM = n$; \quad $\fdM \parallel m_{Si} \gets \fdM$\,
    \State \Assert $\mathsf{MAC}(k_{Si}, ct) = m_{Si}$
    \State $m_{iR} \gets \mathsf{MAC}(k_{iR}, ct)$; \quad $\fdM \gets m_{iR} \parallel \fdM$
    
    \State $h{ct} \gets \mathsf{H}(ct)$; \quad $\mathcal{P}^\prime \gets \mathcal{P} \cup \{hct\}$
    \State \begin{varwidth}[t]{\linewidth}
        $\mathcal{S}_i \gets \mathcal{S}_i \setminus \{(sid, \mathtt{OK}, n, i, N_{i - 1}, N_{i + 1}, N_{i+2}, k_{Si}, k_{iR}, \tau_{i-1}, r_i, \mathcal{P})\}$ \par
        \hskip5mm ${} \cup \{(sid, \mathtt{OK}, n, i, N_{i - 1}, N_{i + 1}, N_{i+2}, k_{Si}, k_{iR}, \tau_{i-1}, r_i, \mathcal{P}^\prime)\}$
        \end{varwidth}
    \State $\fdsrc \gets N_i$; \quad $\fddst \gets N_{i + 1}$; \quad $\fdidx \gets i$
    \State Transmit $(h \parallel p)$ to $N_{i + 1}$ using the TLS record protocol
\EndProcedure
\vspace{1mm}
\Procedure{dataBackward}{$h \parallel p$}
    \State \Assert $\fdphase = \mathtt{Data} \land \fddir = \mathtt{Backward} \land \fdidx > 0$
    \State $N_{i + 1} \gets \fdsrc$\,; \quad $sid \leftarrow \fdsid$\,; \quad $i \gets \fdidx$
    \State \begin{varwidth}[t]{\linewidth}
        \Assert $\exists n, N_{i-1}, N_{i+2}, k_{Si}, k_{iR}, \tau_{i-1}, r_i, \mathcal{P}$: \par
        \hskip5mm $(sid, \mathtt{OK}, n, i, N_{i - 1}, N_{i + 1}, N_{i+2}, k_{Si}, k_{iR}, \tau_{i-1}, r_i, \mathcal{P}) \in \mathcal{S}_i$
        \end{varwidth}
    \State $\fdsrc \gets N_i$; \quad $\fddst \gets N_{i - 1}$; \quad $\fdidx \gets i - 1$
    \State Transmit $(h \parallel p)$ to $N_{i + 1}$ using the TLS record protocol
\EndProcedure
    \end{algorithmic}
  \end{algorithm}

  \begin{algorithm}[t]
    \scriptsize
    \captionsetup{font=small}
    \caption{~Receiver $R$ in the data transmission phase}
    \label{alg:data_receiver}
    \begin{algorithmic}
\Procedure{dataForward}{$h \parallel p$}
    \State \Assert $\fdphase = \mathtt{Data} \land \fddir = \mathtt{Forward}$
    \State $N_n \gets \fdsrc$\,; \quad $sid \gets \fdsid$\,; \quad $n \gets \fdidx$\,; \quad $ct \gets p$

    \State \begin{varwidth}[t]{\linewidth}
        \Assert $\exists ts, k_{SR}, k_{1R}, ..., k_{nR}, \tau_{n}, x_0G, \sigma_S, K, C, \Pi$: \par
        \hskip1.5mm $(sid, \mathtt{OK},  n, n + 1, N_{n}, ts, k_{SR}, k_{1R}, ..., k_{nR}, \tau_{n}, x_0G, \sigma_S, K, C, \Pi) \in \mathcal{S}_{n+1}$
        \end{varwidth}

    \State \Assert $\#\fdM = n$; \quad $m_{nR} \parallel ... \parallel m_{1R} \gets \fdM$
    \For{$i = n$ \textbf{downto} $1$}
        \State \Assert $\mathsf{MAC}(k_{iR}, ct) = m_{iR}$
    \EndFor
    
    \State $pt \gets \mathsf{KDec}(k_{SR}, ct)$
    \If{$\mathscr{C}_R(ts, pt) = 0$}
        \State Invoke \textsc{report}$(sid, pt, ct)$
    \Else
        \State Deliver $(sid, pt)$ to the upper layer
    \EndIf
\EndProcedure
\vspace{1mm}
\Procedure{dataBackward}{$sid, pt$}
    \State \Assert $\fdphase = \mathtt{Data} \land \fddir = \mathtt{Backward}$
    \State \begin{varwidth}[t]{\linewidth}
        \Assert $\exists n, N_n, ts, k_{SR}, k_{1R}, ..., k_{nR}, \tau_{n}, x_0G, \sigma_S, K, C, \Pi$: \par
        \hskip1.5mm $(sid, \mathtt{OK}, n, n + 1, N_{n}, ts, k_{SR}, k_{1R}, ..., k_{nR}, \tau_{n}, x_0G, \sigma_S, K, C, \Pi) \in \mathcal{S}_{n+1}$
        \end{varwidth}
    \State $ct \gets \mathsf{Enc}(k_{SR}, pt)$
    \State Initialize $h$; \quad $\fdsrc \gets R$; \quad $\fddst \gets N_n$; \quad $\fdsid \gets sid$
    \State $\fdidx \gets n$; \quad $\fdphase \gets \mathtt{Data}$; \quad $\fddir \gets \mathtt{Backward}$
    \State Transmit $(h \parallel ct)$ to $N_1$ using the TLS record protocol
\EndProcedure
    \end{algorithmic}
  \end{algorithm}

\subsection{Path Establishment Phase}

To communicate with a receiver $R$, a sender $S$ selects $n$ relays $N_1, ..., N_n$, randomly or strategically~\cite{barton2018towards,rochet2020claps}, using information from the public directory.
While $n = 3$ suffices for relationship anonymity, $n \geq 5$ ensures that the relay learning $S$ and those learning $R$ remain unaware of one another; this is analogous to Tor's three relays design, but \pname requires two more relays because each relay also learns its two-hop-ahead successor. \looseness=-1

$S$ then executes the \textsc{pathForward} procedure in Algorithm~\ref{alg:path_sender}.
It generates an ECDH key pair $(x_0, x_0G)$, derives the session identifier $sid = \mathsf{H}(sid)$, and produces a traceable signature $\sigma_S$ over $x_0G$ and a timestamp $ts$.
$x_0$ is combined with each on-path node's long-term public key $pk_i$ to derive the session key $k_{Si}$, which is used to encrypt the list of \emph{encrypted forwarding information} $E$.
Through $E$, $S$ securely informs each on-path relay or the receiver of the path length, its position, and its immediate predecessor, successor, and two-hop-ahead successor.
Subsequently, it initializes a key list $K$ and a chain of successor proofs $(C, \Pi)$.
It also computes the predecessor proof $\tau_0$, the successor proof $\pi_0$ to be inserted into $\Pi$, and the corresponding confirmation $\rho_0$.
The values $sid$, $x_0G$, $ts$, $\sigma_S$, $E$, $K$, $(C, \Pi)$, $\tau_0$, and $\rho_0$, are then packed into a packet and sent to $N_1$.
The packet transfer between $S$ and $N_1$, as well as all subsequent interactions between adjacent on-path nodes, occurs over secure channels realized by TLS connections between them. \looseness=-1

Upon receiving this packet from $N_{i-1}$, $N_i$ ($1 \leq i \leq n$) executes the \textsc{pathForward} procedure in Algorithm~\ref{alg:path_relay}.
It first derives $k_{Si}$ from $sk_i$ and $x_0G$ to decrypt the respective forwarding information in $E$.
It also verifies the received $\tau_{i-1}$ and $\rho_{i-1}$, generates its own ECDH key pair $(x_i, x_iG)$, and inserts $x_iG$ into $K$.
Next, it computes the predecessor proof $\tau_i$, the commitment $c_i$ to be inserted into $C$, the successor proof $\pi_i$ to be inserted into $\Pi$, and its confirmation $\rho_i$.
Finally, it forwards the packet, now containing the updated $K$, $(C, \Pi)$, $\tau_i$, and $\rho_i$, to $N_{i+1}$.

Once the packet reaches $R$, it executes the \textsc{pathReply} procedure in Algorithm~\ref{alg:path_receiver}. 
It verifies $\sigma_S$, confirms that no known trapdoor can trace it, and then verifies $\tau_n$ and $\rho_n$.
Since $(C, \Pi)$ now encompasses the entire path, $R$ stores it together with $K$ for the violation report phase.
It then generates an ECDH key pair $(x_{n+1}, x_{n+1}G)$, derives the session key $k_{SR}$ and the signature $\sigma_R$ through one-way authenticated key exchange using $x_0G$, and derives $k_{iR}$ for each $N_i$ through ECDH using $x_iG$ in $K$.
Finally, it replies to $N_n$ with a packet containing $x_{n+1}G$ and $\sigma_R$.

On the return path of this packet, each $N_i$ executes the \textsc{pathBackward} procedure in Algorithm~\ref{alg:path_relay} to derive $k_{iR}$. When the packet finally returns to $S$, it derives $k_{SR}$ and verifies $\sigma_R$ in \textsc{pathBackward} procedure in Algorithm~\ref{alg:path_sender}.

\subsection{Data Transmission Phase}

In the forward direction, $S$ runs \textsc{dataForward} procedure in Algorithm~\ref{alg:data_sender}.
It encrypts the plaintext $pt$ into $ct$ with $k_{SR}$ using key-committing AE, computes $M$, a list of MACs $m_{Si}$ on $ct$ with $k_{Si}$ for each $N_i$, and sends a packet containing $M$ and $ct$ to $N_1$. \looseness=-1

Upon receiving this packet, each relay $N_i$ executes \textsc{dataForward} procedure in Algorithm~\ref{alg:data_relay}.
It verifies $m_{Si}$ using $k_{Si}$, computes $m_{iR}$ on $ct$ with $k_{iR}$, and insert it into $M$.
The packet is forwarded to $N_{i+1}$, and $hct$, the hash of $ct$, is recorded at $N_i$.
To reduce the memory required to store $hct$, receivers may be required to report packets for $T$ seconds (e.g., $T = 86400$) after session establishment.
This is practical given that contracts are immediately evaluable.
Alternatively, allowing a constant error enables efficient $hct$ storage using probabilistic data structures. \looseness=-1

Once the packet reaches $R$, it executes \textsc{dataForward} procedure in Algorithm~\ref{alg:data_receiver}.
It verifies $m_{iR}$ with $k_{iR}$ for each $N_i$, decrypts $ct$, and checks that $pt$ complies with the contract at $ts$. 

In the backward direction, packets are simply forwarded in reverse under end-to-end encryption: $R$ encrypts messages, each $N_i$ forwards the ciphertext, and $S$ decrypts it.

\subsection{Violation Report Phase}

  \begin{algorithm}[t]
    \scriptsize
    \captionsetup{font=small}
    \caption{~Receiver $R$ in the violation report phase}
    \label{alg:report_receiver}
    \begin{algorithmic}
\Procedure{report}{$sid, pt, ct$}
    \State \begin{varwidth}[t]{\linewidth}
        \Assert $\exists n, N_n, ts, k_{SR}, k_{1R}, ..., k_{nR}, \tau_{n}, x_0G, \sigma_S, K, C, \Pi$: \par
        \hskip1.1mm $(sid, \mathtt{OK}, n, n + 1, N_{n}, ts, k_{SR}, k_{1R}, ..., k_{nR}, \tau_{n}, x_0G, \sigma_S, K, C, \Pi) \in \mathcal{S}_{n+1}$
        \end{varwidth}

    \State \begin{varwidth}[t]{\linewidth}
        $\mathcal{S}_{n + 1} \gets \mathcal{S}_{n + 1} \setminus \{(sid, \mathtt{OK}, n, n + 1, N_{n}, ts, k_{SR}, k_{1R}, ..., k_{nR},$ \par
        \hskip19mm $ \tau_n, x_0G, \sigma_S, K, C, \Pi)\} \cup \{ (sid, \mathtt{Shut}) \}$
        \end{varwidth}
        
    \State Send $(R, pt, ct, n, N_n, ts, k_{SR}, \tau_n, x_0G, \sigma_S, K, C, \Pi)$ to $V$ securely
\EndProcedure
\vspace{2mm}
\Procedure{reportAccepted}{$sid, td_S$}
    \State \Assert $td_S \neq \bot \land \exists (sid, \mathtt{Shut}) \in \mathcal{S}_{n+1}$
    \State $\mathcal{S}_{n+1} \gets \mathcal{S}_{n+1} \setminus \{(sid, \mathtt{Shut})\} \cup \{(sid, \mathtt{Trapdoor}, td_S)\}$
\EndProcedure
    \end{algorithmic}
  \end{algorithm}

  \begin{algorithm}[t]
    \scriptsize
    \captionsetup{font=small}
    \caption{~Verifier $V$ in the violation report phase}
    \label{alg:report_verifier}
    \begin{algorithmic}
\Procedure{reportReceived}{$R, pt, ct, n, N_n, ts, k_{SR}, \tau_n, x_0G, \sigma_S, K, C, \Pi$}
    \State \Assert The report is received from $R$
    \State $sid \gets \mathsf{H}(x_0G)$
    \State \textbf{if} {\begin{varwidth}[t]{\linewidth}
        $\#K \neq n \lor \#C \neq n \lor \#\Pi \neq n + 1 \lor \mathscr{C}_R(ts, pt) = 1 \lor {}$ \par
        $\mathsf{KDec}(k_{SR}, ct) \neq pt \lor \mathsf{Verify}(pk_n^\prime, sid \parallel R \parallel \bot, \tau_n) = 0\lor {}$ \par
        $\mathsf{TVerify}(gpk, x_0G \parallel ts, \sigma_S) = 0$ \textbf{then}
        \end{varwidth}} 
        
        \State \hskip\algorithmicindent \Return $(sid, \bot)$ to $R$, and blame $R$ for the invalid report;
    \State \textbf{end if}
        
    \State $hct \gets \mathsf{H}(ct)$; \quad $N_n^\prime \gets N_n$; \quad $N_{n+1}^\prime \gets R$
    \For{$i = n$ \textbf{downto} $1$}

        \State \begin{varwidth}[t]{\linewidth}
            Send query $(sid, K, C, \Pi, hct)$ to $N_i^\prime$ securely and \par
            \hskip11mm get response $(sid, \bar{\upsilon}_{i}, \upsilon_{i}, r_i, \tau_{i-1}, N_{i-1}^\prime, b_i)$
        \end{varwidth}
        
        \State $\hat{K} \parallel x_iG \gets K$; \quad $\hat{C} \parallel c_i \gets C$; \quad $\hat{\Pi} \parallel \pi_i \gets \Pi$
        \If{$\mathsf{UVerifyD}(pk_{i}^\prime, sid \parallel K \parallel C \parallel \hat{\Pi}, \pi_i, \bar{\upsilon}_i) = 1$}
            \State \Return $(sid, \bot)$ to $R$, and blame $N_{i + 1}^\prime$ for the trace disruption; 
        \State \hskip-\algorithmicindent\textbf{else if} {\begin{varwidth}[t]{\linewidth}
        $\mathsf{UVerifyC}(pk_{i}^\prime, sid \parallel K \parallel C \parallel \hat{\Pi}, \pi_i, \upsilon_i) = 0 \lor \mathsf{Com}(r_i, \tau_{i-1}) \neq c_i \lor {}$ \par
        $\mathsf{Verify}(pk_{i-1}^\prime, sid \parallel N_{i}^\prime \parallel N_{i+1}^\prime, \tau_{i-1}) = 0$ \textbf{then}
        \end{varwidth}} 
            \State \Return $(sid, \bot)$ to $R$, and blame $N_{i}^\prime$ for the trace disruption; 
        \EndIf
        \State $K \gets \hat{K}$; \quad $C \gets \hat{C}$; \quad $\Pi \gets \hat{\Pi}$
    \EndFor
    \State $S \gets \mathsf{TOpen}(gsk, \sigma_S)$
    \If{$N_0^\prime \neq S$}
        \State \Return $(sid, \bot)$ to $R$, and blame $N_0^\prime$ for the trace diversion;
    \ElsIf{ $b_1 + ... + b_n \leq n/2$ }
        \State \Return $(sid, \bot)$ to $R$, and blame $R$ for the invalid report; 
    \Else
        \State $td_S \gets \mathsf{TReveal}(gsk, S)$; 
        \State \Return $(sid, td_S, S)$ to $R$, and blame $S$ for the malicious message $pt$
    \EndIf
\EndProcedure
    \end{algorithmic}
  \end{algorithm}

  \begin{algorithm}[t]
    \scriptsize
    \captionsetup{font=small}
    \caption{~Relay $N_i$ in the violation report phase}
    \label{alg:report_relay}
    \begin{algorithmic}
\Procedure{reportQueried}{$sid, K, C, \Pi, hct$}
    \State \begin{varwidth}[t]{\linewidth}
        \Assert $\exists n, i, N_{i-1}, N_{i+1}, N_{i+2}, k_{Si}, k_{iR}, \tau_{i-1}, r_i, \mathcal{P}$: \par
        \hskip5mm $1 \leq i \leq n \land (sid, \mathtt{OK}, n, i, N_{i - 1}, N_{i + 1}, N_{i+2}, k_{Si}, k_{iR}, \tau_{i-1}, r_i, \mathcal{P}) \in \mathcal{S}_i$
        \end{varwidth}
    \State $\hat{\Pi} \parallel \pi_{i} \gets \Pi$; \quad
    $\bar{\upsilon}_{i} \gets \mathsf{UDisavow}(sk_i, sid \parallel K \parallel C \parallel \hat{\Pi}, \pi_i)$ 
    \State $\upsilon_{i} \gets \mathsf{UConfirm}(sk_i, sid \parallel K \parallel C \parallel \hat{\Pi}, \pi_{i})$; \quad $b_i \gets \left|\mathcal{P} \cap \{hct\}\right|$ 
    
    \State Send $(sid, \bar{\upsilon}_{i}, \upsilon_{i}, r_i, \tau_{i-1}, N_{i-1}, b_i)$ to $V$ securely
\EndProcedure
    \end{algorithmic}
  \end{algorithm}

When $R$ detects a message plaintext $pt$ violating the contract, it runs the \textsc{report} procedure in Algorithm~\ref{alg:report_receiver}: submit $pt$, $ct$, $k_{SR}$, $x_0G$, $ts$, $\sigma_S$, $n$, $K$, $(C, \Pi)$, $\tau_n$, and the identity of $N_n$ to $V$.

Upon receiving the report, $V$ executes the \textsc{reportReceived} procedure in Algorithm~\ref{alg:report_verifier}.
It first verifies that $pt$ indeed violates the receiver's contract at time $ts$, that $ct$ correctly decrypts to $pt$ under $k_{SR}$, and the validity of $\sigma_S$ and $\tau_n$. 
Any failure in these verifications indicates an invalid report, in which case $V$ holds $R$ accountable. 
Next, $V$ begins verifying $(C, \Pi)$ from $N_n$ by querying each $N_i$ about the traversal of $ct$; each relay responds by executing the \textsc{reportQueried} procedure in Algorithm~\ref{alg:report_relay}.
If any failure occurs during this process, $V$ holds the corresponding relay accountable. 
If fewer than a majority of the $n$ relays respond affirmatively about the traversal of $ct$, $V$ holds $R$ accountable.
$V$ then opens $\sigma_S$, and if the opened identity, say $S$, differs from the that identified from successful verification of $(C, \Pi)$, say $N_0^\prime$, differs from $S$, $V$ holds $N_0^\prime$ accountable.
Only when all of the above conditions are satisfied does $V$ provide $R$ with the trapdoor of $S$ and hold $S$ accountable.

Upon obtaining the trapdoor for $S$ from $V$, $R$ stores it to shut off future path establishment from $S$.

%% file: security.tex
\section{Security Analysis}
\label{sec:security}

\subsection{Anonymity}
\subsubsection{Outside eavesdropper}
No relationship anonymity is leaked to non-participating parties along the path. 
Since all messages between \pname parties are protected by TLS, the adversary learns nothing beyond the identities of the immediate previous and next-hop nodes. 
Session unlinkability is also guaranteed, as outside parties cannot observe sessions.

\subsubsection{Malicious relay}
No relationship anonymity or session unlinkability is leaked to a single relay, as knowledge revealed to it is limited to the path length, its own position on the path, and the identities of its immediate predecessor, the successor, and the successor two hops ahead.
$e_{i}$ informs the relay $N_i$ of the path length, its position,  predecessor, successor, and two-hop ahead successor.
$\tau_{i-1}$ and $\rho_{i-1}$ reveal nothing beyond the predecessor and successor. 
Entries of $E$ are encrypted, and $(C, \Pi)$ leaks only negligible knowledge, according to Theorem~\ref{thm:succ_proof_chain}.
All other observed values including $x_0G$, $\sigma_S$, $x_{n+1}G$, $\sigma_R$, $K$, and $M$ are also computationally independent of anonymity.

\subsubsection{Malicious receiver}
No anonymity is leaked to the receiver unless it succeeds in reporting the sender.
The discussion follows from that for a malicious relay.
The message plaintext $pt$ is also learned, but leakage from the upper layer is out of scope.


\subsection{Accountability}

\subsubsection{Abuse by sender/relays}
A malicious sender may send malicious messages, and colluding relays may disrupt the report.
However, the verifier can identify the sender or the last misbehaving relay under the honest majority assumption. 

\vspace{0.3mm}
\draftparagraph{Repudiation of session-sender links} 
A sender transmitting malicious messages cannot evade accountability.
It must submit a valid traceable signature in the session establishment phase, allowing the verifier to identify the sender later by the unforgeability of the traceable signature scheme.

An adversary compromising multiple senders may submit a traceable signature under an identity different from the actual message sender revealed by the path-session link.
In this case, \pname holds the malicious message sender accountable.

\vspace{0.3mm}
\draftparagraph{Repudiation of path-session links}
In the path establishment phase, a malicious relay may miscompute or tamper with the chain of successor proofs.
However, any such attempt either causes this relay to fail to submit a confirmation or prompts its predecessor to issue a disavowal in the violation report phase, thereby rendering the relay accountable according to Theorem~\ref{thm:chain_nonrep}.
In the violation report phase, a relay cannot halt the report without being blamed unless it provides a valid disavowal accusing its successor. 
Additionally, an honest relay cannot be incorrectly blamed, as guaranteed by Theorem~\ref{thm:chain_framing}. \looseness=-1

\vspace{0.3mm}
\draftparagraph{Repudiation of packet-path links}
In the violation report phase, a malicious relay may falsely deny a packet traversal.
However, the verifier's judgment remains correct according to Theorem~\ref{thm:data_all}, under the honest-majority assumption.

\vspace{0.3mm}
\draftparagraph{Repudiation of plaintext-packet links}
In the data transmission phase, a malicious sender may craft a message that causes decryption failure at the verifier.
However, this also causes decryption failure at the receiver, and the message is discarded.

\subsubsection{Path trace diversion}
\label{sec:misdirection}
In the violation report phase, malicious relays may divert reconstruction of a path established by a malicious sender onto another maliciously established path.
However, participant accountability is still guaranteed. 

Consider the following attack: A malicious sender $S$ starts establishing a path to a honest receiver $R$, $\mathsf{Path}=(S, ..., N_{i-1}$, $N_{i}$, $N_{i+1}$, $N_{i+2}, ..., R)$, with session identifier $sid$, where $N_i$ colludes with $S$. 
$N_i$ intercepts this path establishment and induces a colluding off-path node $S^\prime$ to construct another path $\mathsf{Path}^\prime=(S^\prime, ..., N_{i-1}^\prime$, $N_{i}$, $N_{i+1}$, $N_{i+2}^\prime, ..., R^\prime)$, where $N_{i-1}^\prime \neq N_{i-1}$, using the same $sid$. 
$N_i$ then reuses the predecessor proofs of $N_{i-1}^\prime$ on $\mathsf{Path}^\prime$ to compute its successor proofs on $\mathsf{Path}$, thereby diverting the verifier's tracing from $\mathsf{Path}$ onto $\mathsf{Path}^\prime$ at $N_i$. \looseness=-1

As noted before Theorem~\ref{thm:chain_branch}, the verifier does not detect this diversion at $N_i$. 
Instead, it continues tracing $\mathsf{Path}^\prime$ backward from $N_i$ until some relay is held accountable. 
Or, if the verifier finally reaches $S^\prime$, which differs from $S$, the original sender opened by the traceable signature, and holds $S^\prime$ accountable for constructing the malicious path instead of $S$.

\subsubsection{Framing by relays/receiver}
\label{sec:framing}
A malicious receiver and a colluding relay may attempt to frame the sender to obtain her trapdoor for de-anonymization.
However, such attempts are detected by the verifier under the honest majority assumption. \looseness=-1

\vspace{0.5mm}
\draftparagraph{Retroactive contracts}
A receiver can update the contract at any time, but the updated contract never applies to sessions established before the update. 
Any retroactive report based on the updated contract fails, since the sender's traceable signature is bound to the timestamp at the path establishment.

\vspace{0.5mm}
\draftparagraph{Forgery of session-sender links}
A malicious receiver cannot produce a traceable signature that opens to an honest sender due to the framing resistance of the traceable signature scheme. \looseness=-1

\vspace{0.5mm}
\draftparagraph{Forgery of path-session links}
A malicious receiver and a colluding relay may attempt to forge a chain of successor proofs or to divert the verifier's reconstruction of a maliciously established path onto an honestly established path. 
However, such forgery is detected, and framing an honest sender through such a diversion is impossible, due to Theorems~\ref{thm:succ_proof_chain} and \ref{thm:chain_branch}.

\vspace{0.5mm}
\draftparagraph{Forgery of packet-path links}
In the violation report phase, a malicious relay falsely confirms the traversal of a packet.
However, the verifier's judgment remains correct under the honest majority assumption, according to Theorem~\ref{thm:data_all}.

\vspace{0.5mm}
\draftparagraph{Forgery of plaintext-packet links}
In the violation report phase, given a received ciphertext, a malicious receiver submits an arbitrarily chosen plaintext and session key to the verifier. 
However, malformed plaintexts are rejected due to key-committing AE. \looseness=-1


\subsection{Non-goals and Limitations}
The following properties are non-goals of \pname because they inherently conflict with its primary goals.

\subsubsection{Resisting colluding eavesdroppers}
The local adversary assumption is essential because packet-path linkage requires relays to recognize identical packets, thereby precluding bitwise unlinkability. Consequently, anonymity is preserved against adversaries compromising only a partial segment of the path (e.g., small-scale ISPs), but not against adversaries observing both ends of the path (e.g., state-level surveillance authorities).

\subsubsection{Security without honest majority}
The honest majority assumption is essential for packet-path linkage and is precisely the tradeoff for eliminating per-packet signatures.
Without per-packet signatures, if at least half of the $n$ on-path relays can be Byzantine, the verifier cannot distinguish the case where one half is Byzantine from the case where the other half is Byzantine.

\subsubsection{No trusted verifiers}
Eliminating the verifiers (whether centralized or distributed as in the extension discussed in Section~\ref{sec:discussion}) significantly complicates the protocol, as such a trustee is expected to exercise several privileges, including (i) decrypting ciphertexts to determine whether they violate the contract; (ii) opening the sender's anonymous credential; and (iii) ultimately imposing penalties on identified misbehaving parties outside the protocol.
The first can be eliminated by requiring the receiver to directly prove contract violations to the relays in zero knowledge, albeit at considerable cost.
Democratic group signatures~\cite{manulis2006linkable,zheng2008democratic} may replace the second, but doing so requires consensus among the group members. 
Eliminating the third requires even stronger consensus to enforce penalties. \looseness=-1

\subsubsection{Universal contracts}
We deliberately restrict contracts to well-defined predicates to prevent arbitrary interpretation as a means of de-anonymizing senders.
This restriction is not merely a limitation of our protocol, but reflects the fundamental distinction between disputes that are computationally tractable, and those that are not. 
For example, whether ``This program halts'' constitutes misinformation is generally undecidable.

%% file: evaluation.tex
\section{Performance Evaluation}
\label{sec:perf}

\begin{figure}[t]
\centering
\begin{minipage}[b]{0.70\linewidth}
\vspace{0mm}
\adjustimage{width=1\linewidth,left}{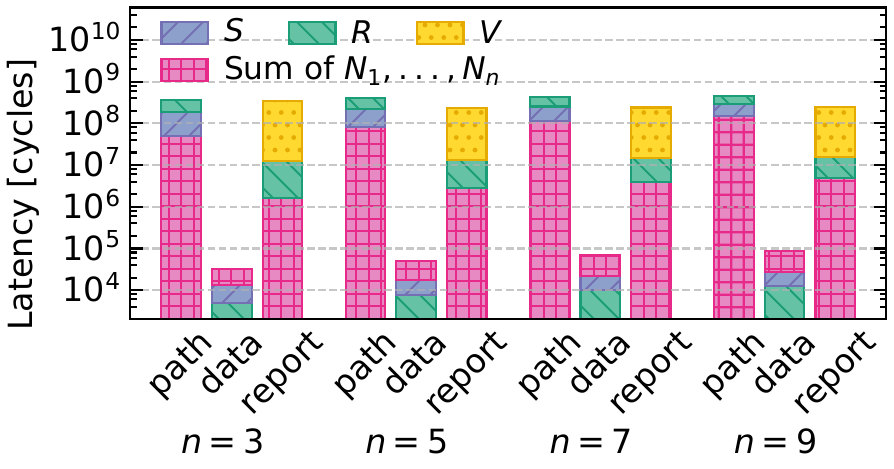}
\vspace{-5mm}
\subcaption{End-to-end latency in each phase}
\label{fig:alllatency}
\end{minipage}
\\ \vskip2.5mm
\begin{minipage}[b]{0.59\linewidth}
\vspace{0mm}
    \begin{minipage}[b]{0.52\linewidth}
    \vspace{0mm}
    \adjustimage{width=1\linewidth,left}{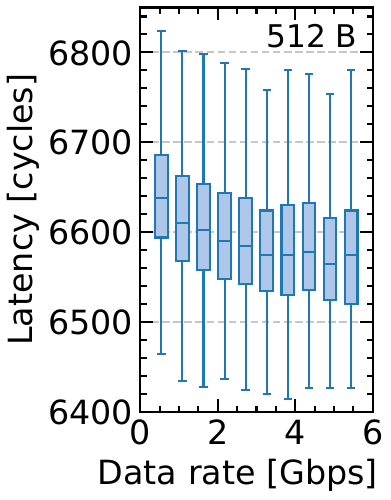}
    \end{minipage}
    \begin{minipage}[b]{0.462\linewidth}
    \vspace{0mm}
    \adjustimage{width=1\linewidth,left}{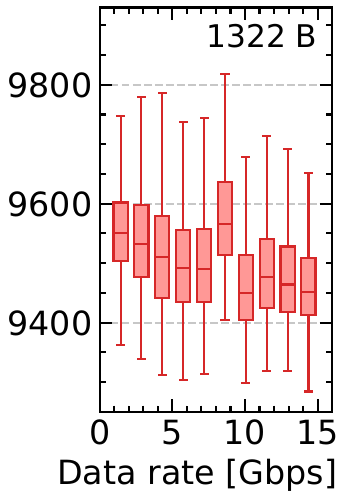}
    \end{minipage}
\subcaption{Relay latency}
\label{fig:latency}
\end{minipage}
%
%
\begin{minipage}[b]{0.36\linewidth}
\vspace{0mm}
\adjustimage{width=1\linewidth,left}{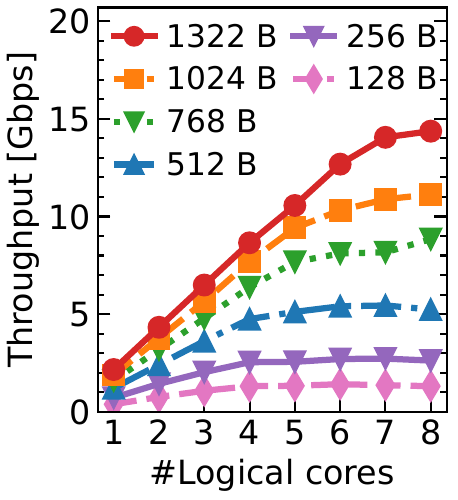}
\vspace{1mm}
\vspace{-5mm}
\subcaption{Relay throughput}
\label{fig:throughput}
\end{minipage}
\vspace{-0.5mm}
\caption{Evaluation result}
\label{fig:result}
\end{figure}

\subsection{Experimental Setup}
We implement a \pname sender, relay, receiver, and verifier on a computer equipped with an Intel Xeon Gold 6330 CPU and an Intel E810 NIC. 
The relay is implemented using DPDK~\cite{dpdk}.

We use Ed25519 for signatures, X25519 for ECDH key exchange, and Curve25519 for one-way authenticated key exchange and undeniable signatures.
AES-256-GCM and GMAC are used for encryption and MAC, and SHA-256 is used as the cryptographic hash function. 
Key-committing AE is constructed from AES-256-GCM, and commitment is derived from SHA-256 in the random oracle model.
The above primitives are implemented using OpenSSL~\cite{openssl}. 
For traceable signatures, we use libgroupsig~\cite{diaz2015libgroupsig} configured with a 3072-bit RSA modulus. 

\subsection{Computation and Communication}
\draftparagraph{Path establishment/data transmission}
Figure~\ref{fig:alllatency} presents the mean total latency, measured in clock cycles over $10^4$ trials, incurred by each node in each phase.
A data packet is configured to convey data whose size corresponds to 512 bytes in plaintext.
The result shows that processing in the data transmission phase is approximately four orders of magnitude lighter than the other phases, highlighting the \emph{dataplane efficiency} achieved by eliminating per-packet public-key cryptography.
For example, in the case of $n=3$, a data packet incurs around $10^4$ cycles at each node.
In contrast, in the path establishment phase, each $N_i$ incurs around $10^7$ cycles, primarily due to the undeniable signatures, while $S$ and $R$ incur around $10^8$ cycles due to traceable signatures.
Even so, these values are still reasonable for the slow path of a cryptography-intensive protocol~\cite{naous2011verifying}.

Figure~\ref{fig:latency} presents the latency distributions of a relay in the data transmission phase, each over $10^4$ packets, under varying data forwarding rates from 10\% to 100\% of the throughput, with $n=3$ and 512- or 1322-byte data.
The latency remains consistently below $10^4$ cycles regardless of the data rate.
These values include TLS encryption and decryption.
Figure~\ref{fig:throughput} presents the relay throughput in the data transmission phase, with $n=3$, a varying number of utilized CPU logical cores, and data sizes between 128 and 1322 bytes.
With 8 cores and 1322-byte data, the throughput reaches 14.37~Gbps. 

\vspace{0.5mm}
\draftparagraph{Violation report}
The violation report phase requires around $10^6$, $10^7$, and $10^8$ cycles at $N_i$, $R$, and $V$, respectively.
However, the reporting performance cannot be characterized solely by these baseline clock-cycle counts:
First, the cost of evaluating the contract on the plaintext at both $R$ and $V$ is not included, as it depends on the specifics of the contract.
Second, because $V$ interacts with each relay $N_i$ a total of $n$ times, the overall reporting latency is dominated by the sum of the $n$ RTTs.

\subsection{Memory Space}
The state size at $N_i$ determines the maximum number of concurrent flows that can be served.
Specifically, our implementation maintains 132 bytes of state per session for the session's lifetime and a 32-byte hash per packet for $T$ seconds. 
The former must reside in DRAM because it is accessed in the data transmission phase, whereas the latter is accessed only in the violation report phase and can therefore be offloaded to HDDs. 

The former state is comparable to that of ordinary network functions, whereas the latter state is the primary concern.
Assume $T = 86400$ seconds and that a 32-byte hash is recorded for each 512-byte packet arriving at a throughput of 5.4~Gbps, the upper bound shown in the above evaluation. 
This requires 3.6~TB of space and a sustained write bandwidth of 42~MB/s; this is within the capacity of commodity server-grade HDDs.

%% file: discussion.tex
\section{Potential Extensions}
\label{sec:discussion}
This section discusses potential directions for extending \pname in terms of anonymity and accountability.

\vspace{0.8mm}
\draftparagraph{Distributing verifiers}
Assuming a central, trusted verifier may be undesirable from the perspective of deploying an anonymity system in practice. 
The following extension distributes the trusted role across $m$ verifiers, thereby relaxing the assumption to requiring only that at least $t_V$ ($>m/2$) of them are honest. 

We replace traceable signatures with \emph{group signatures with $(t_V,m)$-threshold traceability}~\cite{ghadafi2014efficient,blomer2015short,gennaro2019fully}. 
Upon receiving a report from a receiver, each verifier checks the contract violation and the correctness of the decryption. 
Only if at least $t_V$ verifiers find the report valid, they cooperatively open the sender's signature using the threshold traceability functionality.
Similarly, each relay responds to a verifier's query only after receiving it from at least $t_V$ distinct verifiers.
Thus, the sender is identified only by the cooperation of at least $t_V$ verifiers; otherwise, they learn nothing about the sender's identity. \looseness=-1

\vspace{0.8mm}
\draftparagraph{Broader contracts}
The definition of contracts in Section~\ref{sec:contract} may seem too restrictive.
One extension is to let the verifier(s), rather than a publicly evaluable function, decide whether a message is malicious, similar to consensus-based or subjectivity-based contracts~\cite{jonathancontractual}.
This improves expressiveness, but makes it harder to hold receivers accountable for framing reports.

\vspace{0.8mm}
\draftparagraph{Presumption of innocence}
Although \pname already prevents framing by receivers, the principle of \emph{presumption of innocence} may demand stricter justification before accusing the sender.
We can raise the quorum $t_N$ required to conclude a packet-path link from query responses of $n$ relays, from a majority ($t_N =\lfloor n/2 \rfloor+1$) to, e.g., unanimity ($t_N=n$).
This introduces a trade-off, making it harder to prosecute senders colluding with relays. \looseness=-1

\vspace{0.8mm}
\draftparagraph{Reputation systems} Enabling receivers to ensure the honest majority assumption remains an open problem.
A possible approach is a reputation system~\cite{resnick2000reputation,das2014re3}, in which each relay's record of judgments on previously confirmed violations is made public. Relays and receivers assign lower trust to relays with shorter track records. Nevertheless, holding a malicious sender accountable when it colludes with a majority of on-path relays with well-established track records is beyond the capability of a protocol without per-packet signatures. 

\vspace{0.8mm}
\draftparagraph{Beyond bitwise adversary}
Website fingerprinting~\cite{panchenko2011website} and certain traffic analysis attacks~\cite{murdoch2005low} can be mitigated by integrating orthogonal techniques, such as dummy packets~\cite{dyer2012peek}. 
In contrast, attacks by multiple on-path corruptions, such as traffic confirmation, fall outside the local-adversary assumption.

\vspace{0.8mm}
\draftparagraph{Denial-of-service mitigation}
A malicious sender in \pname can issue a large number of path establishment requests, forcing relays to perform expensive undeniable-signature operations and undermining anonymity~\cite{borisov2007denial}.
Such an attack can be mitigated by introducing a client puzzle based on proof-of-work~\cite{juels1999client}.

%% file: conclusion.tex
\section{Conclusion}
This paper proposed \pname, a protocol that reconciles anonymity and accountability.
Through security analysis, implementation, and performance evaluation, we demonstrated that this reconciliation is realistic.